  \documentclass[12pt]{article}
  \usepackage{amsmath}
  \usepackage{verbatim}
  \usepackage{textcomp}
  \usepackage{amssymb, latexsym, amsmath}
  \usepackage[dvips]{color}
  \usepackage{graphicx}
  \usepackage{color}
  \usepackage{ifpdf}
  \usepackage[latin1]{inputenc}
  \usepackage{amsfonts}
  \usepackage{amsbsy}
  \usepackage{amstext}
  \usepackage{amsopn}
  \usepackage{amsgen}
  \usepackage{natbib}
  \newcommand{\be}{\begin{equation}}
  \newcommand{\ee}{\end{equation}}
  \newcommand{\ba}{\begin{eqnarray}}
  \newcommand{\ea}{\end{eqnarray}}
  \newcommand{\bs}{\begin{subequations}}
  \newcommand{\es}{\end{subequations}}
  \newcommand{\Omegabold}{\mbox{{\boldmath $\Omega$}}}
  
  \newcommand{\s}{\nobreak\hspace{.11em}\nobreak}

 \title{Relaxation of Viscoelastic Tumblers,\\
  with Application to 1I/2017 (`Oumuamua) and 4179 Toutatis}

\author{
 James A. Kwiecinski\\
 \small{Mathematics, Mechanics, and Materials Unit,}\\
 \small{Okinawa Institute of Science and Technology, Okinawa 904-0495, Japan}\\
 \small{james.kwiecinski{\s}@{\s}oist.jp}\\
  ~\\
  }
     \date{}

\begin{document}

\maketitle

 \begin{abstract}
 Motivated by the observation of comets and asteroids rotating in non-principal axis (NPA) states, we investigate the relaxation of a freely precessing
 triaxial ellipsoidal rotator towards its lowest-energy spin state. 
 Relaxation of the precession arises from internal dissipative stresses generated by self-gravitation and inertial
 forces from spin. We develop a general theory to determine the viscoelastic stresses in the rotator, under any linear rheology, for both
 long-axis (LAM) and short-axis (SAM) modes. By the methods of continuum mechanics,
 we calculate the power dissipated by the stress field and the viscoelastic material strain which enables us to determine the timescale of the precession dampening. To illustrate
 how the theory is used, we apply our framework to a triaxial 1I/2017 (\textquoteleft Oumuamua) and 4179 Toutatis under the Maxwell regime. For the former, employing viscoelastic parameters typical of very cold monolithic asteroids renders a dampening timescale longer by a factor of $10^{10}$ and
 higher than the timescales found in the works relying on the $\,Q$-factor approach, whilst the latter yields a significantly shorter timescale as a consequence of including self-gravitation. We further reduce our triaxial theory to bodies of an oblate geometry and derive a family of relatively simple analytic approximations determining the NPA dampening times for Maxwell rotators, as well as a criterion determining whether self-gravitation is negligible in the relaxation process. Our approximations exhibit a relative error no larger than $0.2\%$, when compared to numerical integration, for close to non-dissipative bodies and $0.002\%$ for highly energy dissipating rotators.
 \end{abstract}




 \section{Preliminaries}

 \subsection{Tumbling comets and asteroids}

 Precessing unsupported tops are not uncommon in astronomy. Pulsars, planets, comets, asteroids, and cosmic dust granules often tumble, i.e., rotate in NPA
 (non-principal-axis) states. In the Light Curve Database, NPA characteristics are exhibited by 497 out of 19640 objects for which the rotation has been measured reliably \citep[updated on 31 January 2019]{warner}.
 Most of these tumblers are of small to medium size, with their diameters seldom exceeding $20$ km \citep[Figure 8]{Pravec_et_al_2014_(99942)_Apophis}. 
 
 Examples of tumbling comets include P/Halley \citep{sagdeev1989rotation}, 46P/Wirtanen \citep{samarasinha1996comments,rickman1998comet}, 
 29P/Schwachmann-Wachmann~1 \citep{meech1993nucleus}, and 67P/Churyumov-Gerasimenko \citep{gutierrez2016possible}. Among asteroids, notable examples of tumblers are 
 4179~Toutatis \citep{Ostro}, 2008 TC3 \citep{Scheirich}, and 99942~Apophis \citep{Pravec_et_al_2014_(99942)_Apophis}. Recent work has also suggested that the interstellar asteroid 
 1I/2017~U1 (`Oumuamua) was rotating in an NPA state during its fly-by past Sol \citep{drahus2018tumbling,fraser2018tumbling,belton2018excited,rafikov2018,Bannister2019}.

 There exist many physical mechanisms by which tumbling can be excited, such as gravitational torques \citep{kwiecinski2018effects},
 outgassing \citep{jewitt1997measurements}, collisions with other celestial objects \citep{henych2013asteroid}, the YORP effect
 \citep{breiter2015tumbling}, or formation of the tumbling body through disruption of a progenitor \citep{giblin}.

 The inertial forces emerging in a tumbling rotator contain oscillating components, which consequently result in periodic stresses. As no rotator is perfectly elastic, these stresses cause internal friction, which entails energy dissipation, without affecting the angular
 momentum. Once excited, a free rotator evolves towards a spin state corresponding to a minimal energy, with a fixed value of the angular momentum
 vector; the state of rotation around the shortest principal axis (which is the axis with the maximal moment of inertia). This end-state is achieved
 in the situations where an external factor excites tumbling and then becomes negligible, so that free rotation is allowed to occur. A more complex system is a setting where ongoing external excitation is competing with dissipation, however, this latter setting is beyond the scope of our paper.

 \subsection{History and recent progress}

 Heretofore, in almost all studies on the topic, the dissipation rate was parameterised with an empirical quality factor $\,Q\,$.
 Within this approach, the following estimate of the damping time was offered by \cite{burns1973asteroid}:
 \ba
 \tau
 \,\propto~\frac{\mu~Q}{\rho\,R^2\,\Omega^3}~A~~,
 \label{1}
 \ea
 with $\,\mu\,$, $\,Q\,$, $\,\rho\,$, $\,R\,$ being the mean shear rigidity, quality factor, density, and radius of the body; $\,\Omega\,$ being the
 spin rate. In their formulae (22 - 23), the authors estimated the numerical factor $\,A\,$ to be about a hundred for near-spheroidal rotators:
 $\,A^{\textstyle{^{(Burns~et~al)}}}\,\propto~100\,$.

 A milestone result, this estimate was, however, very approximate and, as we know now, rendered an inflated value for $\,A\,$. More importantly, the
 estimate did not provide $\,A\,$ as a function of a residual nutation angle. These shortcomings motivated several authors to improve the calculation, such as solving a boundary-value problem for stresses and strains, and employing the resulting solutions in a subsequent calculation of the energy dissipation
 rate that depended on the nutation angle.

 For oblate bodies, such analysis was suggested by \citet{efroimsky2000inelastic}. Those authors noticed that a large part of dissipation in an oblate
 rotator comes from the second harmonic, which is a double of the precession frequency.$\,$\footnote{~The second harmonic emerges due to the centrifugal
 force being quadratic in the angular velocity $\,\Omegabold\,$. For a dynamically oblate rotator, the components of $\,\Omegabold\,$ are proportional
 to  $\,\sin\omega\,t\,$ and $\,\cos\omega\,t\,$, where $\,\omega\,$ is the nutation rate and $\,t\,$ is time. Hence, in the expression for the centrifugal
 force, squaring of $\,\Omegabold\,$ gives birth to $\,\sin 2\omega\,t\,$ and $\,\cos 2\omega\,t\,$ terms. Such terms then emerge in the stress and
 strain tensors, thereby affecting the dissipation rate.

In triaxial rotators, precession generates stresses at an infinite number of frequencies $\,\chi_n\,$ which are overtones of some $\,${\it{base frequency}}
  $\,\chi_1\,$ that is lower than $\,\omega\,$ (see Section \ref{sec:Nome} for details).
 \label{kelen}} This was one of the reasons for those authors obtaining a much faster relaxation rate: $~A^{\textstyle{^{(E\&L)}}}\,\propto~1 - 4~$.
 The authors modeled the body with a rectangular prism, and the boundary conditions for the stresses were satisfied on its surfaces only approximately.

 \citet{molina2003energy} applied their method to an oblate ellipsoid. They too imposed the boundary conditions approximately, leading to the values $~A^{\textstyle{^{(Molina~et~al)}}}\,\propto~10 - 30~$.

 \citet{sharma2005nutational} solved the equation for displacements, with exact boundary conditions. For oblate bodies, they obtained
 $~A^{\textstyle{^{(Sharma~et~al)}}}\,\propto~200 - 800~$ and found even larger values for prolate shapes which yielded very long timescales of relaxation.

 \citet{breiter2012stress} developed an exact solution for displacements in an elastic triaxial ellipsoid. These authors found that, while excessively large
 values of $\,A\,$ were obtained by \citet{sharma2005nutational} due to the accumulation of three mathematical oversights, overtly small values were obtained by
 \citet{efroimsky2000inelastic} (by a factor of $\,14/\pi\,$) mainly due to modeling the body with a prism of a volume higher than any solid of revolution with the
 same ratio of axes.

 In all those works, calculations comprised two main steps: First, the stress and strain tensors (or the field of displacements) were found under the
 assumption that the body was elastic. Second, the elastic energy was calculated and an empirical quality factor $\,Q\,$ was introduced to account for the
 energy damping rate. This rate was then used to calculate the decay rate of the precession cone.

The two aforementioned steps are, however, incompatible. On the one hand, elasticity implies instantaneous reaction, i.e., a zero
 phase lag between the deformation and stressing whilst, on the other hand, calculation of the power damped at a certain frequency yields a quality factor whose
 inverse is equal to the sine of the phase lag at this frequency \cite[Appendix A]{frouard2017precession}. The error caused by employment of this method
 will increase with increasing deformability of the material. Specifically, the method is inapplicable to those asteroids and comets which are rubble. In
 such bodies, lagging between action and reaction forces is large and the effective viscosity becomes an important parameter that must enter the
 calculation of deformation caused by the precession-generated stressing \citep{efr2015}. The calculation should be based on a rheological law, and should
 render the phase lag at each frequency of the deformation spectrum. These lags should then enter the calculation of the energy dissipation rate at each
 frequency. From this rate, it is then possible to calculate the precession dampening timescale as a function of the half-angle of the precession cone.

 In a recently published paper by \cite{frouard2017precession}, this process was followed for oblate ellipsoids, using elastic stresses from \cite{sharma2005nutational}
 and corresponding these to viscoelastic stresses in Fourier space for a linear Maxwell rheology. The scope of the current work, whilst having the same goal,
 seeks to extend this previous work to triaxial ellipsoids using a different, more mechanically general formalism. We aim to derive the linear elastic stresses in a triaxial
 ellipsoidal rotator, using a direct, stress-based approach built on the framework of \cite{breiter2012stress}; and to derive the corresponding viscoelastic
 stresses in the Laplace space, using Residue Theory which we argue is more algebraically feasible for complicated geometries with fewer axes of
 symmetry. We further discuss subtleties involving the kinematics of a freely rotating triaxial rotator, such as the different rotational behaviours it can
 exhibit, like the long-axis (LAM) and short-axis (SAM) modes, and the analytical difficulties in modeling the transition between them, as well as pursuing a mechanical treatment of the problem.

 To make our historical account complete, we would mention a fully numerical approach to the problem, recently suggested by \citet{quillen2019simulations}.
 That method can be employed as an independent test for analytical models.

 \subsection{Plan of the paper}

 We organise the paper as follows: In Section 2, we describe the kinematics and mechanics of a rotating object, without the effects of torques.
 We explain that, in the infinitesimal deformation regime which is the focus of the work, one can treat the rotator as quasi-rigid and solve
 the Euler equations to determine the rotational behaviour over short timescales. 
 
 In Section 3, we formulate a framework to determine the elastic
 stresses emerging in a homogeneous ellipsoid due to its rotation and self-gravitation, whilst in Section 4 we employ the Correspondence
 Principle in Laplace space to obtain the viscoelastic stresses under an arbitrary linear rheology. In Section 5, from our knowledge of the
 stress field and rheology, we determine the rate of energy dissipation and then calculate the timescale necessary to dampen the precession angle.

To illustrate our theory in practice, we apply it to a triaxial `Oumuamua- and Toutatis-sized object obeying the Maxwell rheology and discuss the role of parameters, such as mass density and aspect ratio, in the relaxation process in Section 6. We further reduce our general triaxial theory to bodies of an oblate geometry to facilitate comparison with previous work and subsequently derive analytic approximations for the dampening timescale, as well as a criterion for when self-gravitation can be ignored, in Section 7. We conclude the present work with a discussion of our results.

 \section{Freely rotating quasi-rigid ellipsoids\label{label}}

 We assume that the unperturbed (no-wobble) shape of the body is not very different from a triaxial ellipsoid and was acquired by the body long
 ago in the course of its accretion. Indeed, dependent on the rotation rate, the figures of a stable equilibrium of an inviscid fluid can be
 either oblate ({\it{a Maclaurin ellipsoid}}) or triaxial ({\it{Jacobi ellipsoid}}). Other shapes are available but unstable, see \citet{grigor}.
 We also assume that the body, once shaped, has the capability to sustain its geometry. This implies that the body, even if highly porous and cracked,
 is stronger than rubble and has enough bonds to retain its shape after the minimal energy state is reached and the stress becomes stationary. Thus,
 whatever viscoelastic model we use, it will be applicable to minor variations of shape only during precession relaxation. Mind, though, that to solve for these small displacements,
 we shall need to know both the oscillating and constant components of the stress.

 \subsection{Preliminary theory: Quasi-rigid approximation}\label{sec:preliminary}

 We consider a homogeneous rotator of an ellipsoidal geometry with mass $\,m\,$. To describe its dynamics, we use a basis comprising the unit vectors
 $\boldsymbol{e}_{1}$, $\boldsymbol{e}_{2}$, and $\boldsymbol{e}_{3}$ that move with the body and always align with its principal axes.
 The lengths of the ellipsoid's semi-major axes are $a$, $b$, and $c$ and we further define their ratios
 \footnote{~To facilitate comparison of our formalism with that developed for oblate rotators in \citet{frouard2017precession}, we note that their parameter
 $\,h\,$ coincides with our $\,h_2\,$. There is no $\,h_1\,$ because oblateness implies $\,a=b\geq c\,$ and $\,h_1=1\,$.}
 \begin{align}
 h_{1} = & \frac{b}{a}, \qquad h_{2} =  \frac{c}{b}\,\;.
 \label{ratios}
 \end{align}
 In this co-rotating basis, a position vector $\boldsymbol{r}$ of a small parcel of material is given by
 \begin{equation}
 \boldsymbol{r}=x\boldsymbol{e}_{1}+y\boldsymbol{e}_{2}+z\boldsymbol{e}_{3}\,\;, \label{eq:position}
 \end{equation}
whilst its instantaneous angular velocity $\boldsymbol{\Omega}$ in the rotating frame is
 \begin{equation}
 \boldsymbol{\Omega}=\Omega_{1}\boldsymbol{e}_{1}+\Omega_{2}\boldsymbol{e}_{2}+\Omega_{3}\boldsymbol{e}_{3}\,\;.
 \end{equation}

 As the co-rotating basis is set to always align with the principal axes, the inertia tensor $\,\mathbb{I}\,$  always stays diagonal and time-independent
 \begin{equation}
 \mathbb{I}=\left[\begin{array}{ccc}
 \frac{\textstyle ma^{2}}{\textstyle 5}h_{1}^{2}\left(1+h_{2}^{2}\right) & 0 & 0\\
 0 & \frac{\textstyle ma^{2}}{\textstyle 5}\left(1+h_{1}^{2}h_{2}^{2}\right) & 0\\
 0 & 0 & \frac{\textstyle ma^{2}}{\textstyle 5}\left(1+h_{1}^{2}\right)
 \end{array}\right]\;\;. \label{eq:I}
 \end{equation}
 Conservation of the angular momentum in the co-rotating frame renders the Euler equations of motion for a free top
 \begin{equation}
 \frac{\mathrm{d}}{\mathrm{d}t}\left(\mathbb{I}\boldsymbol{\Omega}\right)=\boldsymbol{\Omega}\times\mathbb{I}\boldsymbol{\Omega}\,\;.
 \end{equation}
 The time derivative on the left-hand side may be interpreted as the rate of change of the angular momentum vector in the rotating reference frame while the right-hand side is related to the moment of inertial forces \citep{landau1976mechanics}.

 The essence of the quasi-rigid approximation is the decoupling of the precession dynamics, which occurs on the short timescale, and the body deformation that occurs on the large timescale. As a result, we suppose the body keeps its shape as it tumbles, which implies that
 $\,\dot{\mathbb{I}} \, \boldsymbol{\Omega} \,\ll\,\mathbb{I} \; {\dot{\boldsymbol{\Omega}}}_i\;$; so the equations of rotational motion become
 \begin{equation}
 \mathbb{I}\frac{\mathrm{d}\boldsymbol{\Omega}}{\mathrm{d}t}\approx\boldsymbol{\Omega}\times\mathbb{I}\boldsymbol{\Omega}\,~. \label{eq:rigid}
 \end{equation}

 \subsection{Integrals of motion and regimes of rotation}

\begin{figure}
 \begin{center}

   \includegraphics[width=0.45\textwidth]{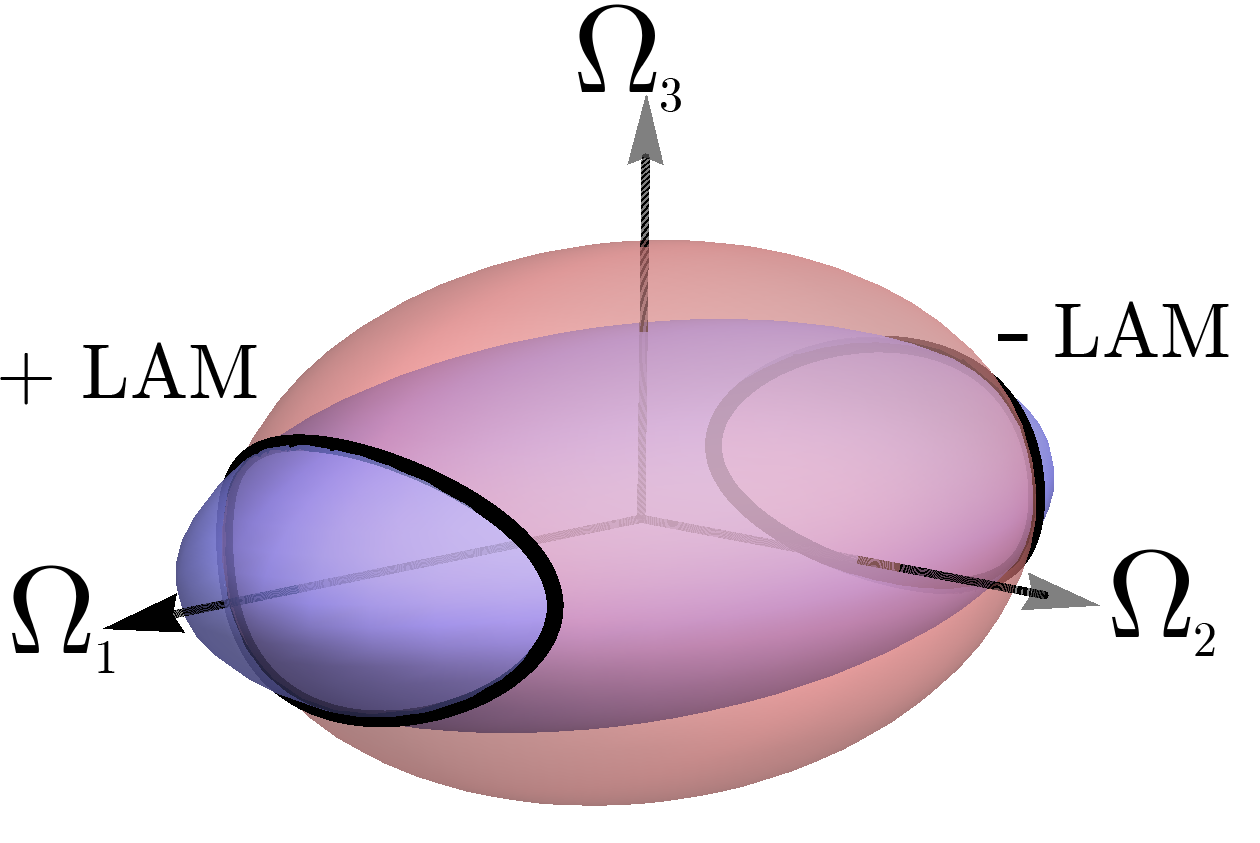}\\
   \includegraphics[width=0.45\textwidth]{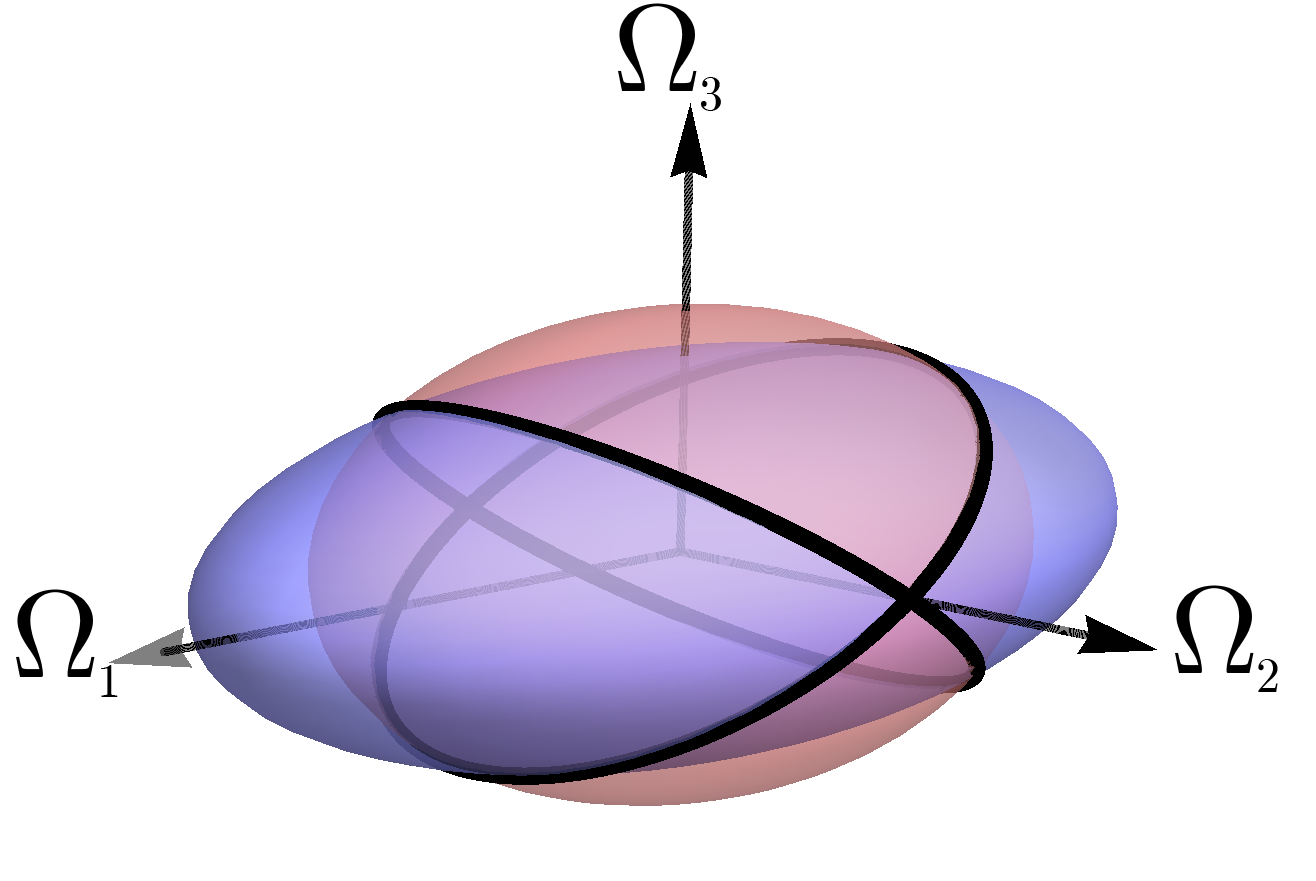}\\
   \includegraphics[width=0.45\textwidth]{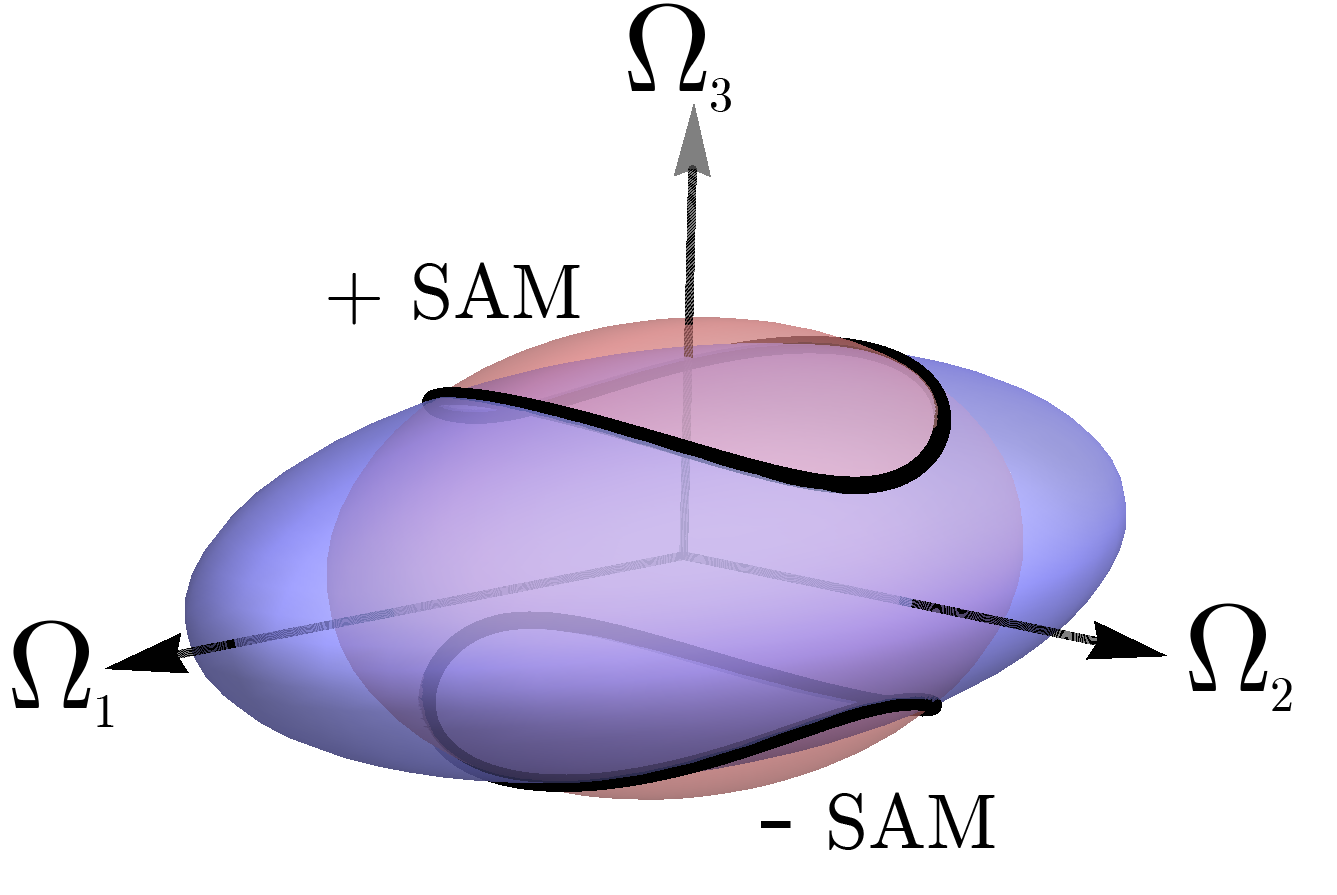}
 
 \end{center}
 \caption{Illustration of precession relaxation, in the space of body-frame angular velocities, for a quasi-rigid ellipsoid with principal
 semi-major axes lengths $a>b>c$. Red and blue ellipsoids are the surfaces of constant $T_{kin}$ and $\left|\boldsymbol{J}\right|$,
 as specified in \eqref{eq:Tkin} and \eqref{eq:J}, whilst the black trajectories are the intersection curves and therefore solution trajectories of
 \eqref{eq:rigid}. Large kinetic energies give precession solutions around the minimal inertial axis in LAM rotation (top), with the two solutions
 corresponding to the clockwise and anti-clockwise orientations. Such solutions are quantified by a maximum wobbling angle $\theta_{max}^{\left(L\right)}$
 defined in \eqref{eq:thetaLAM}. As the kinetic energy decreases, these two solution curves intersect to form the separatrix (middle) where
 $\theta_{max}^{\left(L\right)}=90^{\circ}$ and, for lower kinetic energies than this state, the rotation now precesses around the maximal inertial axis
 in SAM rotation (bottom) which is quantified by the wobbling angle $\theta_{max}^{\left(S\right)}$ defined in \eqref{eq:thetaSAM}. (Color online) \label{fig:SAMLAM}}
 \end{figure}

Equation \eqref{eq:rigid} possesses two integrals of motion~---~the kinetic energy $\,T_{kin}\,$ and the magnitude of the
 angular momentum vector $\,\left|\boldsymbol{J}\right|\,$. With $\,I_{ij}\,$ being the elements of the matrix
 $\,\mathbb{I}\,$ given by \eqref{eq:I}, the conservation of the kinetic energy and the magnitude of the angular momentum reads
 \begin{align}
 2T_{kin}\,= & \;I_{11}\Omega_{1}^{2}+I_{22}\Omega_{2}^{2}+I_{33}\Omega_{3}^{2}\,\;,\label{eq:Tkin}\\
 \nonumber\\
 {{\bf{J}}^{2}}\,= & \;I_{11}^{2}\Omega_{1}^{2}+I_{22}^{2}\Omega_{2}^{2}+I_{33}^{2}\Omega_{3}^{2} \label{eq:J}\,\;.
 \end{align}
 Mathematically, in the space of the body-frame angular velocities $\,\Omega_i\,$, each solution $\,\Omegabold(t)\,$ to the Euler equations is characterised
 by fixed values of $\,T_{kin}\,$ and $\,{\bf{J}}\;$. This situation is illustrated by Figure \ref{fig:SAMLAM} where the red and blue ellipsoids are the surfaces
 of constant $\,T_{kin}\,$ and $\,\left|{\bf{J}}\right|\,$, correspondingly. For a fixed value of $\,|{\bf{J}}|\,$, three different values of $\,T_{kin}\,$ are
 considered.

 A solution to the Euler equations (\ref{eq:rigid}) coincides with the moving tip of a
 vector $\,\boldsymbol{\Omega}(t)\,$ pointing from the origin to a point on the surface of the angular-momentum ellipsoid. The tip describes a trajectory made by the
 intersection of the kinetic-energy ellipsoid with the angular-momentum ellipsoid~---~a thick black line in the figure. We see that two distinct classes of
 solutions are possible: When the angular-velocity vector is closer to the minimal-inertia axis
 and precesses around it, the corresponding solution is termed a $\,${\it{Long-Axis Mode}}$\,$ (LAM) whilst if the angular-velocity vector is closer
 to the maximal-inertia axis and precesses about it, the solution is called a $\,${\it{Short-Axis Mode}}$\,$ (SAM). The curve (in fact, a union of two intersecting
 curves) dividing the two classes is the $\,${\it{separatrix}}.

 In one extreme case, the kinetic-energy ellipsoid exhibiting the largest value of $\,T_{\small{kin}}\,$ available for a fixed $\,|{\bf{J}}|\,$ would embed the angular-momentum ellipsoid, touching it in two opposite points on the $\,\Omega_{1}\,$ axis. These points
 correspond to rotation about the minimal-inertia axis or the longest principal axis. In the other extreme case, the kinetic-energy ellipsoid
with the smallest available value of $\,T_{\small{kin}}\,$ would be located inside the angular-momentum ellipsoid, and would be
 touching it from inside in two opposite points on the $\,\Omega_{3}\,$ axis. These points are the states of complete relaxation of precession or the rotation about the shortest principal axis. In both extreme cases, we obtain two solutions
 corresponding to two possible orientations of the spin~---~clockwise and counter-clockwise.

 \subsection{Measure of precession and the adiabatic approximation \label{sec:precession}}

 Aside from these extreme cases, the spin mode is NPA and the rotator precesses. To quantify the sweep of precession, consider the angle made by the angular
 momentum $\,{\bf{J}}\,$ and the body axis about which the angular momentum is precessing
 \begin{align}
 \mbox{LAM}\,:\qquad\quad
 \theta^{(L)}\,\equiv\,\arccos\left(\frac{{\bf{J}}\cdot{\bf{e}}_{1}}{|{\bf{J}}|}\right)\,\;,
 \label{}\\
 \mbox{SAM}\,:\qquad\quad
 \theta^{(S)}\,\equiv\,\arccos\left(\frac{{\bf{J}}\cdot{\bf{e}}_{3}}{|{\bf{J}}|}\right)\,\;.
 \label{}
 \end{align}

 In the simple case of an oblate rotator, only the SAM regime is available and the precession angle $\,\theta\,=\,\theta^{(S)}\,$ does not change on the timescale of the precession
 \citep{efroimsky2000inelastic, efrASR}. In the triaxial case, however, both $\,\theta^{(L)}\,$ and $\,\theta^{(S)}\,$ evolve in time and cannot serve as measures of
 precession. While in \citep{efrJMP} and \citep{efrPSS} it was suggested to measure precession by the time average of $\,\sin^2\theta\,$ over a precession cycle,
 \citet{breiter2012stress} chose to employ the maximal value of the angle over a cycle
 \begin{align}
 \mbox{LAM}\,:\qquad\quad\theta_{max}^{\left(L\right)}=\max\arccos\left(\frac{{\bf{J}}\cdot\boldsymbol{e}_{1}}{\left|{\bf{J}}\right|}\right)\,\;, \label{eq:thetaLAM}\\
 \mbox{SAM}\,:\qquad\quad\theta_{max}^{\left(S\right)}=\max\arccos\left(\frac{\boldsymbol{J}\cdot\boldsymbol{e}_{3}}{\left|\boldsymbol{J}\right|}\right)\,\;.\label{eq:thetaSAM}
 \end{align}
 In our developments hereafter, we shall employ $\,\theta_{max}^{\left(L\right)}\,$ and $\,\theta_{max}^{\left(S\right)}\,$ to facilitate the comparison of our
 results with those of \citet{breiter2012stress}.

 Precession relaxation implies conservation of $\,{\bf{J}}\,$, with a slow decrease of the value of the energy $\,T_{kin}\,$.
 Suppose we have a trajectory that begins in the realm of LAM: With the angular momentum vector precessing about the minimal-inertia (long) axis, the angular velocity vector is moving
 about the axis $\,\Omega_1\,$ in Figure \ref{fig:SAMLAM} top. After some energy is dissipated, the body comes to rotation about the middle-inertia
 axis, with the angular velocity vector pointing to the separatrix, Figure \ref{fig:SAMLAM} middle. This regime is unstable however,
 in that a slight deviation will lead the rotator towards SAM behavior.
 In SAM, the angular momentum vector will be precessing about the maximal-inertia (short) axis, while the angular velocity vector will be spiraling about $\,\Omega_3\,$
 and converging to that axis, see Figure \ref{fig:SAMLAM} bottom.

 In summary, from the maximal energy rotation state, $\,\theta_{max}^{\left(L\right)}\,$ ranges from $\,0^{\circ}\,$ to $\,90^{\circ}\,$, with the latter corresponding
 to the separatrix, given that the solution trajectories intersect the $\Omega_{2}$ axis, therefore yielding an angle of $\,90^{\circ}\,$ when measured with respect to
 the $\,\Omega_{1}\,$ axis. Then, from the separatrix to the minimal energy state, $\,\theta_{max}^{\left(S\right)}\,$ varies from $\,90^{\circ}\,$ to $\,0^{\circ}\,$
 so that the angular momentum vector eventually becomes parallel to the $\,\Omega_{3}\,$ axis.

 Rigorously speaking, dissipation becomes possible only when we return the dropped term
 $\,\frac{\textstyle\mathrm{d\mathbb{I}}}{\textstyle\mathrm{d}t}\boldsymbol{\Omega}\,$ back in the Euler equations. Nonetheless, energy
 decrease may be tolerated when it is adiabatically slow, i.e., when we ensure the separation of timescales
 \begin{align}
 \left\lvert\,
 \frac{d\theta_{max}^{\left(L\right)}}{dt}
 \,\right\rvert
 \;\;\,,\;\;\,
 \left\lvert\,
  \frac{d\theta_{max}^{\left(S\right)}}{dt}\,\right\rvert\;\;\ll\;\;\mbox{precession rate}\;\;\ll\;\;\mbox{rotation rate}\,\;.
 \label{13a}
 \end{align}

 The applicability of the adiabatic approximation depends on the dissipative properties of the material. The validity of the quasi-rigid approximation (that
 $\,\frac{\textstyle\mathrm{d}\mathbb{I}}{\textstyle\mathrm{d}t}\,\boldsymbol{\Omega}\,$ is negligible compared to
 $\,\mathbb{I}\,\frac{\textstyle\mathrm{d}\boldsymbol{\Omega}}{\textstyle\mathrm{d}t}\,$) depends on the deformation properties. To put it roughly, the former
 approximation hinges mainly on viscosity, the latter mainly on elasticity, so we do not expect one of these approximations to entail another, and will treat them as
 independent.

 We would finally mention that when the rotator reaches the minimal energy state the stresses in it become stationary. Under some linear viscoelastic rheologies, this may lead to deformation of the body's axes, however, such processes are outside the scope of the present work. For now, we consider the assumptions made in the
 beginning of Section \ref{label}; namely, that the geometry of the body is unaltered. No matter how close to rubble, the body is assumed to have enough strength to sustain its unperturbed shape.

 \subsection{Solutions for a rigidly rotating triaxial ellipsoid}\label{sec:solutions}

 The solutions to \eqref{eq:rigid} can assume either a LAM or SAM form, dependent on the dimensionless parameter
 \begin{equation}
 \mathcal{B}=\frac{2T_{kin}I_{22}}{\boldsymbol{J}^{2}},
 \end{equation}
 whose values will lie within the interval $\,\frac{\textstyle I_{22}}{\textstyle I_{33}}\leq\mathcal{B}\leq\frac{\textstyle I_{22}}{\textstyle I_{11}}$,
 derived from the fact that the rotational behaviour must be in the region between the minimal energy state
 $\,\boldsymbol{J}^{2}\,=\,I_{33}^{2}\,\Omega_{3}^{2}\,$, $\;2\,T_{kin}\,=\,I_{33}\,\Omega_{3}^{2}\;$ and the maximal energy state
 $\;\boldsymbol{J}^{2}\,=\,I_{11}^{2}\,\Omega_{1}^{2}\,$, $\,2\,T_{kin}\,=\,I_{11}\,\Omega_{1}^{2}\,$.
 When $\,1 < {\mathcal{B}} \leq \frac{\textstyle I_{22}}{\textstyle I_{11}}\,$,
 the rotational motion is of the LAM type, whilst if $\,\frac{\textstyle I_{22}}{\textstyle I_{33}}\,\leq\,{\mathcal{B}}\,<\,1\,$, then SAM rotation occurs.
 Furthermore, given that $\,\mathcal{B}\,$ depends directly on $\,T_{kin}\,$, it is also related to the maximum wobbling angles $\,\theta_{max}^{(L)}\,$
 and $\,\theta_{max}^{(S)}\,$.

 Both the LAM and SAM types of explicit solutions for \eqref{eq:rigid} can be cast elegantly in a single form, which is \citep{deprit1993complete}
 \begin{align}
 \Omega_{1}^{\left(R\right)} & =~\frac{\left|\boldsymbol{J}\right|}{I_{11}}~
         \sqrt{\frac{\left(\mathcal{B}^{\left(R\right)}\,I_{33}-I_{22}\right)I_{11}}{\left(I_{33}-I_{11}\right)I_{22}}}~
         F_{1}^{\left(R\right)}\left(\omega^{\left(R\right)}\left(t-t_{0}\right),k^{\left(R\right)}\right),\\
 \Omega_{2}^{\left(R\right)} & =~\frac{\left|\boldsymbol{J}\right|}{I_{22}}~
         \sqrt{\frac{\mathcal{B}^{\left(R\right)}\,I_{33}-I_{22}}{I_{33}-I_{22}}}~
         F_{2}^{\left(R\right)}\left(\omega^{\left(R\right)}\left(t-t_{0}\right),k^{\left(R\right)}\right),\\
 \Omega_{3}^{\left(R\right)} & =~\frac{\left|\boldsymbol{J}\right|}{I_{33}}~
         \sqrt{\frac{\left(I_{22}\mathcal{\,-~B}^{\left(R\right)}I_{11}\right)I_{33}}{\left(I_{33}-I_{11}\right)I_{22}}}~
         F_{3}^{\left(R\right)}\left(\omega^{\left(R\right)}\left(t-t_{0}\right),k^{\left(R\right)}\right),
 \end{align}
 where $\,t_{0}\in\mathbb{R}\,$ is an arbitrary constant, while the superscript $\,R\,$ can take on two values, $\,L\,$ or $\,S\,$, depending on whether the
 rotational motion is LAM or SAM. Here and hereafter, the dimensionless parameter $\,{\cal{B}}\,$ is denoted with $\,\mathcal{B}^{\left(R\right)}\,$
 where the superscript is needed to determine in which of the two intervals $\,{\cal{B}}\,$ assumes its value. This will be needed below, to emphasise which
 of the two formulae, (\ref{24}) or (\ref{30}), links $\,{\cal{B}}\,$ to the maximal wobble angle.

 Individually, the functions $\,F_{1}^{\left(R\right)},F_{2}^{\left(R\right)},F_{3}^{\left(R\right)}\,$ in LAM rotation are
 \begin{align}
 F_{1}^{\left(L\right)}\left(\omega^{\left(L\right)}\left(t-t_{0}\right),k^{\left(L\right)}\right) & =\pm dn\left(\omega^{\left(L\right)}\left(t-t_{0}\right),k^{\left(L\right)}\right),\\
 F_{2}^{\left(L\right)}\left(\omega^{\left(L\right)}\left(t-t_{0}\right),k^{\left(L\right)}\right) & =k^{\left(L\right)}\mathrm{sn}\left(\omega^{\left(L\right)}\left(t-t_{0}\right),k^{\left(L\right)}\right),\label{eq:addkfactor}\\
 F_{3}^{\left(L\right)}\left(\omega^{\left(L\right)}\left(t-t_{0}\right),k^{\left(L\right)}\right) & =\pm\mathrm{cn}\left(\omega^{\left(L\right)}\left(t-t_{0}\right),k^{\left(L\right)}\right),
 \end{align}
 where the parameter $\omega^{\left(L\right)}$ and the elliptic modulus $k^{\left(L\right)}$  are given by
  \begin{align}
 \omega^{\left(L\right)} & =
 \left|\boldsymbol{J}\right|\sqrt{\left(\frac{1}{I_{11}}~-~\frac{1}{I_{22}}\right)\left(\frac{\mathcal{B}^{\left(L\right)}}{I_{22}}~-~\frac{1}{I_{33}}\right)}\,~,
 \label{22}\\
 k^{\left(L\right)} & =
 \frac{\sin\theta^{\left(L\right)}_{max}}{\sqrt{1~+~\frac{\frac{\textstyle 1}{\textstyle I_{11}}~-~\frac{\textstyle 1}{\textstyle I_{22}}}{\frac{\textstyle 1}{\textstyle I_{22}}~-~\frac{\textstyle 1}{\textstyle I_{33}}}\cos\theta_{max}^{\left(L\right)}}}\,~,
 \label{eq:kLAM}
 \end{align}
 while the maximal wobble angle is related to $\mathcal{B}^{\left(L\right)}$ through \citep{breiter2012stress}
 \begin{eqnarray}
 \mathcal{B}^{\left(L\right)} & =
 ~\textstyle 1 -\left(\textstyle 1-\frac{\textstyle I_{22}}{\textstyle I_{11}}\right)\cos^{2}\theta_{max}^{\left(L\right)}\,~.
 \label{24}
 \end{eqnarray}
 Similarly, in SAM rotation, we have
\begin{align}
F_{1}^{\left(S\right)}\left(\omega^{\left(S\right)}\left(t-t_{0}\right),k^{\left(S\right)}\right) & =\pm\mathrm{cn}\left(\omega^{\left(S\right)}\left(t-t_{0}\right),k^{\left(S\right)}\right)\,~,\\
F_{2}^{\left(S\right)}\left(\omega^{\left(S\right)}\left(t-t_{0}\right),k^{\left(S\right)}\right) & =\mathrm{sn}\left(\omega^{\left(S\right)}\left(t-t_{0}\right),k^{\left(S\right)}\right)\,~,\\
F_{3}^{\left(S\right)}\left(\omega^{\left(S\right)}\left(t-t_{0}\right),k^{\left(S\right)}\right) & =\pm\mathrm{dn}\left(\omega^{\left(S\right)}\left(t-t_{0}\right),k^{\left(S\right)}\right)\,~,
\end{align}
where
\begin{align}
\omega^{\left(S\right)} & =~
\left|\boldsymbol{J}\right|\sqrt{\left(\frac{1}{I_{22}}-\frac{1}{I_{33}}\right)\left(\frac{1}{I_{11}}-\frac{\mathcal{B}^{\left(S\right)}}{I_{22}}\right)}\,~,
\label{28}\\
k^{\left(S\right)} & =
\frac{\sin\theta_{max}^{\left(S\right)}}{\sqrt{1~+~\frac{\frac{\textstyle 1}{\textstyle I_{22}}~-~\frac{\textstyle 1}
{\textstyle I_{33}}}{\frac{\textstyle 1}{\textstyle I_{11}}~-~\frac{\textstyle 1}{\textstyle I_{22}}}~\cos\theta_{max}^{\left(S\right)}}}\,~,
\label{eq:kSAM}
\end{align}
 and the maximal wobble angle is related to $\,{\cal{B}}^{(S)}\,$ as \citep{breiter2012stress}
 \begin{eqnarray}
 \mathcal{B}^{\left(S\right)} & =~\textstyle 1-\left(\textstyle 1-\frac{\textstyle I_{22}}{\textstyle I_{33}}\right)\cos^{2}\theta_{max}^{\left(S\right)}\,~.
 \label{30}
 \end{eqnarray}
 In the above expressions, $\,\mathrm{cn}\left(u,k\right)\,$, $\,\mathrm{sn}\left(u,k\right)\,$, and $\,\mathrm{dn}\left(u,k\right)\,$ are the
 doubly-periodic Jacobi elliptic functions \citep{jacobi1969fundamenta}.

 In summary, the motion of the rotator is described by one constant parameter $\,\left|\boldsymbol{J}\right|$,
 three parameters $\,\left\{I_{11},I_{22},I_{33}\right\}\,$ which are constant under the quasi-rigid approximation,
 and a single time-dependent variable which is the maximum wobbling angle $\theta_{max}^{\left(R\right)}$.
 This angle is related to the rotational kinetic energy of the system, and changes its value as the energy dissipates due to internal
 inelastic stressing.

\section{Elastically deformable triaxial rotators}

 As our eventual aim is calculation of the precession relaxation rate, a key intermediate step will be to find the rate of energy dissipation as a
 function of the precession angle. That rate is equal to the power generated by the internal friction. To compute it, we shall need to know the
 distribution of strains and stresses. While the strain can be found from the stress through the rheological equation for the material, the stress
 is defined by the distribution of the reaction forces in the body.

 \subsection{Forces}

 In an inertial frame, the reaction force per unit mass, $\boldsymbol{f}$, can be found through Newton's Second Law. According to this law, the
 acceleration $\,\boldsymbol{a}\,$ is caused by the combined action of the reaction force $\,\boldsymbol{f}\,$ and the gravity force
 $\,\boldsymbol{b}_{gr}\,$ per unit mass is
 \begin{equation}
-\boldsymbol{f}\,=\,\boldsymbol{a}\,+\,\boldsymbol{b}_{gr}\,\;.
 \label{eq:2ndlaw}
 \end{equation}

 \subsubsection{The acceleration in an inertial frame,\\
  expressed via the body-frame angular velocity}

 The acceleration $\,\boldsymbol{a}\,$ with respect to an inertial frame and the acceleration $\boldsymbol{a}_{rot}$ of a material point in the body frame are linked through the
 expression \citep{goldstein2012klassische,Kwiecinski2019}
 \begin{equation}
 \boldsymbol{a}=\boldsymbol{a}_{rot}+\frac{\mathrm{d}\boldsymbol{\Omega}}{\mathrm{d}t}\times\boldsymbol{r}+
 2\boldsymbol{\Omega}\times\frac{\mathrm{d}\boldsymbol{r}}{\mathrm{d}t}+\boldsymbol{\Omega}\times\left(\boldsymbol{\Omega}\times\boldsymbol{r}\right),
 \end{equation}
 where all time derivatives are rates measured in the rotating basis.

 Rotating with a period $\,\tau \,$, a body of size $\,\mathit{l}\,$ experiences deformation of the order $\,\delta{\mathit{l}}\,\approx\,\epsilon\,{\mathit{l}}\,$.
 In the body frame, parts of the body acquire deformation-caused velocity $\,v\,\approx\,\delta\mathit{l}/\tau\,\approx\,\epsilon\,\mathit{l}/\tau\,$ and
 deformation-caused acceleration $\,a\,'\,\approx\,\delta\mathit{l}/\tau^{2}\,=\,\epsilon\, \mathit{l}/\tau^{2}\,$. These values of $\,v\,$ and $\,a\,'\,$ are much
 smaller than the velocity and acceleration of the body as a whole ($\,{\mathit{l}}/\tau\,$ and $\,{\mathit{l}}/{\tau}^2\,$, correspondingly).
As a consequence, the inertial-frame acceleration reduces to
 \begin{equation}
 \boldsymbol{a}\approx\frac{\mathrm{d}\boldsymbol{\Omega}}{\mathrm{d}t}\times\boldsymbol{r}+\boldsymbol{\Omega}\times\left(\boldsymbol{\Omega}\times\boldsymbol{r}\right)\,~.
 \label{eq:inertial}
 \end{equation}

 In terms of the shape ratios (\ref{ratios}), the moments of inertia are given by (\ref{eq:I}). Insertion thereof in the Euler relations \eqref{eq:rigid} yields the time
 derivative of the angular velocity expressed via its components \citep[Eqn 21]{breiter2012stress}
 \begin{equation}
 \frac{\mathrm{d}\boldsymbol{\Omega}}{\mathrm{d}t}\,=\,-\,\frac{1-h_{2}^{2}}{1+h_{2}^{2}}\,\Omega_{2}\,\Omega_{3}\,\boldsymbol{e}_{1}\,
 +\,\frac{1-h_{1}^{2}\,h_{2}^{2}}{1+h_{1}^{2}\,h_{2}^{2}}
 \,\Omega_{1}\,\Omega_{3}\,\boldsymbol{e}_{2}\,-\,\frac{1-h_{1}^{2}}{1+h_{1}^{2}}\,\Omega_{1}\,\Omega_{2}\,\boldsymbol{e}_{3}\,\;,
 \end{equation}
where the unit vectors $\,\boldsymbol{e}_{1}\,$, $\,\boldsymbol{e}_{2}\,$, and $\,\boldsymbol{e}_{3}\,$ align along the principal axes of inertia.

 \subsubsection{Self-gravitation}

We now consider the contribution to the reaction force by self-gravitation, or the consequence that each material point in the rotator has a mass and will therefore experience
and exert a gravitational force. For a precessing top, the total contribution can be divided into a constant part from the undeformed body $\bar{\boldsymbol{b}}_{gr}$ and an oscillatory part due to the inertial forces from
rotational acceleration $\tilde{\boldsymbol{b}}_{gr}$
 \ba
 {\boldsymbol{b}}_{gr}\,=~\overline{\boldsymbol{b}}_{gr}\,+~{\boldsymbol{\tilde{b}}}_{gr}\,~.
 \label{31}
 \ea

The constant part $\,\overline{\boldsymbol{b}}_{gr}\,$ is the self-gravitation force of an undeformed body. In a point $\,\left(x,\,y,\,z\right)\,$
 inside a \,{\it{homogeneous}}\, ellipsoid parameterized by the body-frame basis $\,\left\{\boldsymbol{e}_{1},\,\boldsymbol{e}_{2},\,\boldsymbol{e}_{3}\right\}\,$,
this contribution is
 \begin{equation}
 \overline{\boldsymbol{b}}_{gr}\,=~-~\gamma_{1}~x~\boldsymbol{e}_{1}~-~\gamma_{2}~y~\boldsymbol{e}_{2}~-~\gamma_{3}~z~\boldsymbol{e}_{3}\,~, \label{eq:bg}
 \end{equation}
 where the constant coefficients $\,\gamma_{1},\,\gamma_{2},\,\gamma_{3}\,$ are \citep{gauss1877theoria, rodrigues1816correspondence}
 \begin{align}
 \gamma_{1} ~=& ~\frac{Gm}{a^{3}}~R_{J}\left(1,h_{1}^{2},h_{1}^{2}h_{2}^{2},1\right)\,~, \label{eq:selfgravstart}\\
 \gamma_{2} ~=& ~\frac{Gm}{a^{3}}~R_{J}\left(1,h_{1}^{2},h_{1}^{2}h_{2}^{2},h_{1}^{2}\right)\,~,\\
 \gamma_{3} ~=& ~\frac{Gm}{a^{3}}~R_{J}\left(1,h_{1}^{2},h_{1}^{2}h_{2}^{2},h_{1}^{2}h_{2}^{2}\right)\,~, \label{eq:selfgravend}
 \end{align}
for $G\,$ being Newton's Gravitational Constant and $\,R_{J}\,$ being an elliptical integral of the form
 \begin{equation}
 R_{J}=\left(u,v,w,p\right)=\frac{3}{2}\intop_{0}^{\infty}\frac{\mathrm{d}s}{\left(p+s\right)\sqrt{\left(u+s\right)\left(v+s\right)\left(w+s\right)}}\,~.
 \end{equation}

 The small oscillating part $\,{\boldsymbol{\tilde{b}}}_{gr}\,$ is due to the periodically evolving distortion caused by the inertial forces. While it
 is not immediately apparent if $\,{\boldsymbol{\tilde{b}}}_{gr}\,$ can be dropped, \citet[Section 3.3.1]{frouard2017precession} provide an argument
 for why the oscillatory part can be neglected for the purpose of calculating the power dissipation. The argument is based on the fact that integrating
 over the volume of the body ultimately averages out this contribution. Within this approximation, we  assume
 \ba
 {\boldsymbol{{b}}}_{gr}\,=\,\overline{\boldsymbol{{b}}}_{gr}\,\;.
 \label{e}
 \ea

 \subsubsection{The force of material reaction}

 By substituting the inertial forces \eqref{eq:inertial} and the effects of self-gravitation \eqref{eq:bg} into Newton's Second Law \eqref{eq:2ndlaw},
 we can write down the reaction force as
 \begin{equation}
 \boldsymbol{f}\,=\;-\;\mathbb{B}\,\boldsymbol{r}\,~,\label{eq:reactionforce}
 \end{equation}
 where the matrix $\,\mathbb{B}\,$ is given by \citep{breiter2012stress}
 \begin{equation}
 \mathbb{B}\,=\left[\begin{array}{ccc}
 \Omega_{2}^{2}\,+\,\Omega_{3}^{2}\,-\,\gamma{}_{1}~~ & ~~ -\;\frac{\textstyle 2\,\Omega_{1}\,\Omega_{2}}{\textstyle 1\,+\,h_{1}^{2}} ~~ & ~~ -\;\frac{\textstyle 2\,\Omega_{1}\,\Omega_{3}}{\textstyle 1\,+\,h_{1}^{2}\,h_{2}^{2}}\\
 -\;\frac{\textstyle 2\,h_{1}^{2}\,\Omega_{1}\,\Omega_{2}}{\textstyle 1\,+\,h_{1}^{2}} & ~~ \Omega_{1}^{2}\,+\,\Omega_{3}^{2}\,-\,\gamma_{2} & -\;\frac{\textstyle 2\,\Omega_{1}\,\Omega_{3}}{\textstyle 1\,+\,h_{2}^{2}}\\
 -\;\frac{\textstyle 2\,h_{1}^{2}\,h_{2}^{2}\,\Omega_{1}\,\Omega_{3}}{\textstyle 1\,+\,h_{1}^{2}\,h_{2}^{2}} & -\;\frac{\textstyle 2\,h_{2}^{2}\Omega_{1}\Omega_{3}}{\textstyle 1+h_{2}^{2}} & ~~~ \Omega_{1}^{2}+\Omega_{2}^{2}-\gamma_{3}
 \end{array}\right]\;\,,
 \label{eq:BB}
 \end{equation}
with its off-diagonal entries related by
 $\,B_{21}\,=\,h_{1}^{2}\,B_{12}\,$, $\,B_{31}\,=\,h_{1}^{2}\,h_{2}^{2}\,B_{13}\,$, and $\,B_{32}\,=\,h_{2}^{2}\,B_{23}\,$.

 \subsection{Determination of linearly elastic stresses}\label{sec:elastic}

 As a first step, we take the body as elastic and calculate its stress field. The solution for elastic stresses found in this section is a necessary step to find the corresponding stresses under a general viscoelastic rheology.

 We have the necessary information to compute the internal elastic stresses of the rotator. Before doing the mathematics, let us make an inventory of
 the assumptions made hereinabove:
 \begin{itemize}
 \item[~1.~] The body is homogeneous and isotropic.
 \item[~2.~] The unperturbed shape of the body is ellipsoidal and was formed long ago. The body is quasi-rigid, in that it preserves its shape over the shortest timescale (rotation about the instantaneous axis), experiences small oscillations of shape over the intermediate timescale (precession) and, again, preserves the average shape over the longest timescale (precession relaxation).
 \item[~3.~] The long-term evolution is adiabatic, in that we treat $\,T_{kin}\,$ as constant over the short and intermediate timescales, and treat it
 as a slowly evolving parameter over the long timescale.
 \item[~4.~] The deformations are small so that we may neglect them in integrations over the volume and, therefore, may use expression (\ref{eq:BB}).
 \end{itemize}
 To these items, we now add:
  \begin{itemize}
 \item[~5.~] The rotator is isolated, with no exterior force or torque acting on it.
 \end{itemize}
 Explicitly or implicitly, these assumptions were employed in the preceding works
 \citep{prendergast1958proceedings,efrASR,sharma2005nutational,molina2003energy,frouard2017precession,breiter2012stress}.
 By items 4 and 5 above, and  by appealing to the balance of the linear and angular momenta \citep{landau1995course},
 we can write down the equations obeyed by the $\,3\times 3\,$ Cauchy stress tensor $\,\boldsymbol{\sigma}\,$  in the body-related frame
 \begin{align}
 \nabla\cdot\boldsymbol{\sigma} & = -\rho\mathbb{B}\boldsymbol{r} \label{eq:BLM}\,~, \\
 \boldsymbol{\sigma} & = \boldsymbol{\sigma}^{T}\,~,
 \label{eq:BAM}
 \end{align}
 where superscript $\,T\,$ denotes transposition, the gradient operator is defined as
 $\,\nabla=\left(\frac{\textstyle\partial}{\textstyle\partial x},\,\frac{\textstyle\partial}{\textstyle\partial y},
 \,\frac{\textstyle\partial}{\textstyle\partial z}\right)\,$, while the mass density $\,\rho\,$ is constant for a homogeneous rotator.
 Furthermore, in expression (\ref{eq:BLM}) we have substituted the reaction force $\,\boldsymbol{f}\,$ with $\,\mathbb{B}\boldsymbol{r}\,$,
 according to \eqref{eq:reactionforce}.

 The balances of linear and angular momentum are not sufficient to completely determine the internal stresses of the body. For example, if one applies a uniaxial load,
 \eqref{eq:BLM} and \eqref{eq:BAM} do not tell us by what measure the material will contract in the perpendicular direction. We must impose a relationship between the
 stress $\boldsymbol{\sigma}$ and the strain of the material $\boldsymbol{\varepsilon}$. Using assumptions 1 and 4, we model our rotator as an homogeneous
 isotropic Hookean body satisfying the constitutive relation
 \begin{equation}
 \boldsymbol{\varepsilon}=\frac{1+\nu}{E}\boldsymbol{\sigma}-\frac{\nu}{E}\mathrm{Tr}\left(\boldsymbol{\sigma}\right)\boldsymbol{I}\,~, \label{eq:Hookean}
 \end{equation}
 where $\nu$ is the Poisson ratio measuring the amount of contraction under a uniaxial load, $E$ is the Young's modulus, $\mathrm{Tr}\left(\square\right)$ is the trace operator,
 and $\boldsymbol{I}$ is the $3\times 3$ identity matrix.

 For the purpose of finding how a body dissipates energy in an NPA state, we only require the calculation of the internal stresses $\boldsymbol{\sigma}$ rather than the complete
 displacement field. To this end, the deformation of the ellipsoid must satisfy the Saint-Venant compatibility condition
\begin{equation}
\nabla\times\nabla\times\boldsymbol{\varepsilon}=0\,~, \label{eq:SaintVernant}
\end{equation}
 which implies that the linear elastic problem always produces a single-valued displacement field, necessary to ensure that a mapping between the deformed and undeformed states
 mathematically exists \citep{slaughter2012linearized}.

 Inserting \eqref{eq:Hookean} in \eqref{eq:SaintVernant} and using the tensor calculus identity
 $$
 \nabla\times\nabla\times\boldsymbol{\varepsilon}=\nabla^{2}\boldsymbol{\varepsilon}+
 \nabla\left(\nabla\mathrm{Tr}\s\boldsymbol{\varepsilon}\right)-
 \nabla\left(\nabla\cdot\boldsymbol{\varepsilon}\right)-\left[\nabla\left(\nabla\cdot\boldsymbol{\varepsilon}\right)\right]^{T}~,
 $$
 we arrive at
\begin{align}
\left(1+\nu\right)\nabla^{2}\boldsymbol{\sigma}+\nabla\left(\nabla\mathrm{Tr}\s\boldsymbol{\sigma}\right)-\nu\nabla^{2}\left(\mathrm{Tr}\s\boldsymbol{\sigma}\right)\boldsymbol{I}\nonumber\\
=\left(1+\nu\right)\left[\nabla\left(\nabla\cdot\boldsymbol{\sigma}\right)+\left[\nabla\left(\nabla\cdot\boldsymbol{\sigma}\right)\right]^{T}\right]\,~.
 \label{expression}
\end{align}
 Given that $\,\nabla\,$ is a spatial differential operator, while the elements of $\,\mathbb{B}\,$ only depend on time, we can rewrite equation \eqref{eq:BLM} as
 $\,\nabla\left(\nabla\cdot\boldsymbol{\sigma}\right)\,=\,-\,\rho\mathbb{B}\,$ and use this to further simplify expression (\ref{expression})
 \begin{equation}
 \left(1+\nu\right)\nabla^{2}\boldsymbol{\sigma}+\nabla\left(\nabla\mathrm{Tr}\s\boldsymbol{\sigma}\right)-\nu\nabla^{2}\left(\mathrm{Tr}\s\boldsymbol{\sigma}\right)\boldsymbol{I}
 \,=\,-\,\rho\left(1+\nu\right)\left[\mathbb{B}+\mathbb{B}^{T}\right]~~. \label{eq:BMequation}
 \end{equation}

 We lastly require a boundary condition to close the system of equations and therefore ensure a unique solution for the elasticity of the rotator.
 To this end, we rely on assumption 2 and also introduce
 \begin{itemize}
 \item[~6.~] No external stresses are applied over the boundary.
 \end{itemize}
 Then, over the boundary $\,\partial B\,$ with unit normal $\,\hat{\boldsymbol{n}}\,$, the Cauchy stress $\,\boldsymbol{\sigma}\,$ satisfies
 \begin{equation}
 \left.\boldsymbol{\sigma}\cdot\hat{\boldsymbol{n}}\right|_{\partial B}=0\,~. \label{eq:BCInt}
 \end{equation}

 The Cartesian coordinates $\,\left(x,y,z\right)\,$ in the co-rotating basis \eqref{eq:position} can be expressed through an ellipsoidal parameterisation. This parameterisation
 comprises the scaled radial coordinate $\,q\in\left[0,1\right]\,$, azimuthal angle $\,\psi\in\left[0,\pi\right]\,$, and polar angle $\,\phi\in\left[0,2\pi\right]\,$
 \begin{eqnarray}
 x   &=& q\;a\;\sin\psi\;\cos\phi\,,
 \label{eq:paramstart}\\
 y   &=& q\;b\;\sin\psi\;\sin\phi\;=\;q\;a\;h_{1}\;\sin\psi\;\sin\phi\,~, \\
 z   &=& q\;c\;\cos\psi ~\qquad~=\;q\;a\;h_{1}\;h_{2}\;\cos\psi   \,~,
 \label{eq:paramend}
 \end{eqnarray}
 where $\,a\,\geq\,b\,>\,c\,$ are the lengths of the ellipsoid's principal axes.

 To determine the unit normal at the boundary of the ellipsoid, we define a function
 \begin{equation}
 f\left(x,y,z\right)~=~\frac{x^{2}}{a^2}\,+\,\frac{y^{2}}{a^{2}\,h_{1}^{2}}\,+\,\frac{z^{2}}{a^{2}\,h_{1}^{2}\,h_{2}^{2}} - 1\,~.
 \end{equation}
 As its gradient is related to the unit normal $\,\hat{\boldsymbol{n}}\,$ through $\,\hat{\boldsymbol{n}}=\nabla f/\left|\nabla f \right|\,$, the boundary
 condition \eqref{eq:BCInt} can be written down as
 \begin{equation}
 \boldsymbol{\sigma}\cdot \left(x\s ,~\frac{y}{h_{1}^{2}}\,,~\frac{z^{2}}{h_{1}^{2}\,h_{2}^{2}}\right)=0\,~,
 \label{eq:BC}
 \end{equation}
 where we employed parameterization (\ref{eq:paramstart} - \ref{eq:paramend}) at $\,q=1\,$. While the lengths of the principal axes $\,\left\{a,\,b,\,c\right\}\,$
 weakly oscillate over the intermediate timescale (precession), the quasi-rigid approximation introduced in Section \ref{sec:preliminary} neglects this effect over
 the short timescale (rotation).

 The rotator's internal stresses are found by solving equations \eqref{eq:BLM}, \eqref{eq:BAM}, and \eqref{eq:BMequation}, subject to the traction-free boundary condition
 \eqref{eq:BC} which have been developed to only depend on the Cauchy stress $\,\boldsymbol{\sigma}\,$. For an ellipsoidal body, we use the ansatz
 \citep{slaughter2012linearized}
 \begin{align}
 \boldsymbol{\sigma}\left(x,y,z,t\right)&=~a^{2}\,\boldsymbol{S}^{\left(00\right)}-\,x^{2}\,\boldsymbol{S}^{\left(11\right)}-\,\frac{y^{2}}{h_{1}^{2}}\,\boldsymbol{S}^{\left(22\right)}
 -\,\frac{z^{2}}{h_{1}^{2}h_{2}^{2}}\,\boldsymbol{S}^{\left(33\right)}\nonumber\\
 &-\,\frac{xy}{h_{1}}\,\boldsymbol{S}^{\left(12\right)}-\,\frac{xz}{h_{1}h_{2}}\,\boldsymbol{S}^{\left(13\right)}-\,\frac{yz}{h_{1}^{2}h_{2}}\,\boldsymbol{S}^{\left(23\right)}~~,
 \label{eq:ansatz}
 \end{align}
where each $\,\boldsymbol{S}^{\left(ij\right)}\,$ is a $\,3\times 3\,$ matrix whose elements are time-dependent only.

 The stress expressed by \eqref{eq:ansatz} has $\,63\,$ unknowns. However, by imposing the balance of angular momentum \eqref{eq:BAM}, the number is reduced to $36$.  Then, the balance
 of linear momentum \eqref{eq:BLM} and the boundary condition \eqref{eq:BC} render us a further $\,30\,$ unique algebraic equations, owing to the relations between the off-diagonal
 entries of $\,\mathbb{B}\,$. One can solve for all the matrix elements in terms of entries of the spatially constant stress $\,\boldsymbol{S}^{\left(00\right)}\,$ and then determine
 the remaining $\,6\,$ entries by solving the constitutive relation \eqref{eq:BMequation}. As the expressions for the $\,\boldsymbol{S}^{\left(ij\right)}\,$ are large and numerous, we
 provide their explicit forms in Appendix \ref{sec:explicit}.

 \section{Rotators obeying linear viscoelastic rheologies}

 We calculate the stresses of a material exhibiting a linear viscoelastic rheology from the elastic stresses found in Section \ref{sec:elastic}. The method which we employ is known
 as the Correspondence Principle. Alternatively named Alfrey-Hoff's Analogy, and sometimes attributed to \citet{biot}, and actually pioneered yet to \citet{darwin}, this approach
 allows us to find intermediate fictitious stresses in an integral transform space where the time-derivative operators become algebraic. Determination of the viscoelastic stresses can then be achieved by inverse transforming these intermediate stresses back into real time. We discuss how these fictitious stresses are derived by a Laplace formalism, and then
 explore how the corresponding viscoelastic stresses can be determined by integrating in the complex plane with Residue Theory.

 \subsection{Correspondence Principle in Laplace Space}

 The isotropic Hookean (elastic) constitutive relation can be decomposed into the deviatoric and volumetric contributions
 \begin{align}
 \boldsymbol{\sigma}_{D}\left(t\right)&=2\mu\boldsymbol{\varepsilon}_{D}\left(t\right), \label{eq:elasticD}\\
 \boldsymbol{\sigma}_{V}\left(t\right)&=3\mathcal{K}\boldsymbol{\varepsilon}_{V}\left(t\right),\label{eq:elasticV}
 \end{align}
 where the subscripts $D$ and $V$ denote the volumetric and deviatoric contributions to the deformation. Specifically, the appropriate components of the stress are
 \begin{align}
 \boldsymbol{\sigma}_{D}&=\boldsymbol{\sigma}-\frac{1}{3}\mathrm{Tr}\left(\boldsymbol{\sigma}\right)\boldsymbol{I},\\
 \boldsymbol{\sigma}_{V}&=\frac{1}{3}\mathrm{Tr
 }\left(\boldsymbol{\sigma}\right)\boldsymbol{I}\,~.
 \end{align}
$\mathcal{K}$ and $\mu$ are the bulk and deviatoric moduli respectively. They are measures of the mechanical resistance to these stresses, and are related to the
 Poisson ratio $\,\nu\,$ by \citep{skrzypek2015mechanics}
 \begin{equation}
 \nu = \frac{3\mathcal{K}-2\mu}{6\mathcal{K}+2\mu}\,~.
 \end{equation}

 Compare the linear elastic problem of \eqref{eq:elasticD} and \eqref{eq:elasticV} with the viscoelastic problem, which can be similarly decomposed into
 deviatoric and volumetric contributions \citep{alfrey1944non,hilton1961extension} \begin{align}
 P_{1}\boldsymbol{\sigma}_{D}\left(t\right)=U_{1}\boldsymbol{\varepsilon}_{D}\left(t\right)\,~,\label{eq:veDt}\\
 P_{2}\boldsymbol{\sigma}_{V}\left(t\right)=U_{2}\boldsymbol{\varepsilon}_{V}\left(t\right)\,~.\label{eq:veVt}
 \end{align}
 Here, $P_{i}$ and $U_{i}$ are linear differential operators in time and are given by
 \begin{align}
 P_{i}&=\sum_{j=0}^{m_{i}}p_{i}^{(j)}\frac{\partial^{j}}{\partial t^{j}},\label{eq:Pop}\\
 U_{i}&=\sum_{l=0}^{n_{i}}u_{i}^{(l)}\frac{\partial^{l}}{\partial t^{l}},\label{eq:Uop}
 \end{align}
 where $p_{i}^{\left(j\right)}$ and $u_{i}^{\left(l\right)}$ are constants, while $i\in\left\{1,2\right\}$, $j\in\left\{0,1,...,m_{i}\right\}$, and $l\in\left\{0,1,...,n_{i}\right\}$.

 By noting the mathematical analogy between the elastic and viscoelastic problems, we introduce a generalised viscoelastic notion of the Poisson ratio
 \begin{equation}
 \nu_{VE} = \frac{P_{1}U_{2}-P_{2}U_{1}}{P_{2}U_{1}+2P_{1}U_{2}}~~.\label{eq:genPR}
 \end{equation}

 We now appeal to the elastic-viscoelastic correspondence principle which provides a framework for determining the stresses of a linear viscoelastic material from the elastic solution.
 The idea is that the constitutive relations \eqref{eq:elasticD} and \eqref{eq:elasticV}, which algebraically relate stresses to strains, map to a fictitious problem in an integral
 transform space where the stress rates and strain rates are similarly algebraic \citep{findley2013creep}. To this end, we introduce the Laplace transform $\mathcal{L}$ which
 maps a function $\,\alpha\left(t\right)\,$ from real time space $\,t\,$ to a function $\,\widehat\alpha(s)\,$ in a complex frequency space $\,s\;$ \citep{pipkin2012lectures}
 \begin{equation}
 \widehat{\alpha}\left(s\right)\equiv\mathcal{L}\left\{\alpha\left(t\right)\right\}=\intop_{0}^{\infty}e^{-st}\alpha\left(t\right)\mathrm{d}t~~, \label{eq:Laplacedef}
 \end{equation}
which features the desirable property of transforming differential operators to algebraic ones \citep{widder2015laplace}
\begin{equation}
\mathcal{L}\left\{\frac{\mathrm{d}^{n}\alpha\left(t\right)}{\mathrm{d}t^{n}}\right\}=s^{n}\hat{\alpha}\left(s\right),\label{eq:Laplaceprop}
\end{equation}
where we have supposed that at some initial time $t=t_{0}$, $\alpha$ and all its derivatives vanish. We emphasize that the forthcoming analysis only applies to rotating bodies which do not exhibit an initial stress and strain as well as initial stress and strain rate.

These aspects allow us to define a corresponding fictitious problem of \eqref{eq:veDt} and \eqref{eq:veVt} in Laplace space
\begin{align}
\widehat{P}_{1}\widehat{\boldsymbol{\sigma}}_{D}\left(s\right)=\widehat{U}_{1}\widehat{\boldsymbol{\varepsilon}}_{D}\left(s\right),\label{eq:veDs}\\
\widehat{P}_{2}\widehat{\boldsymbol{\sigma}}_{V}\left(s\right)=\widehat{U}_{2}\widehat{\boldsymbol{\varepsilon}}_{V}\left(s\right),\label{eq:veVs}
\end{align}
whereby the generalised Poisson ratio \eqref{eq:genPR} is now purely algebraic
\begin{equation}
\widehat{\nu}_{VE} = \frac{\widehat{P}_{1}\widehat{U}_{2}-\widehat{P}_{2}\widehat{U}_{1}}{\widehat{P}_{2}\widehat{U}_{1}+2\widehat{P}_{1}\widehat{U}_{2}}, \label{eq:genPRs}
\end{equation}
according to the property \eqref{eq:Laplaceprop}.

With this framework, one can determine the viscoelastic stresses in $3$ steps: 1. The elastic solution $\boldsymbol{\sigma}\left(\nu,t\right)$ is Laplace transformed to obtain $\widehat{\boldsymbol{\sigma}}\left(\nu,s\right)$, treating $\nu$ as constant. 2. Supposing a material rheology, instances of $\nu$ are replaced with $\widehat{\nu}_{VE}$ as defined in \eqref{eq:genPRs}. 3. The fictitious stress in complex frequency space $\widehat{\boldsymbol{\sigma}}\left(\widehat{\nu}_{VE},s\right)$ is inverse transformed to obtain the viscoelastic stresses in real time $\boldsymbol{\sigma}_{VE}\left(t\right)$.

\subsection{Nome expansions of Jacobi Elliptic Functions}\label{sec:Nome}

 The first step of this process involves Laplace transforming $\boldsymbol{\sigma}$ which, given the forcing from inertial contributions encoded in $\mathbb{B}$ \eqref{eq:BB},
 depends on quadratic terms involving the Jacobi elliptic functions, as defined in Section \ref{sec:solutions}. In general, integrating these functions according to
 \eqref{eq:Laplacedef} cannot be done exactly, so we require the use of Fourier series called Nome expansions to determine the fictitious stresses analytically.
 To obtain expressions for the mixed terms, we employ the standard Nome expansions \citep{byrd2013handbook} for $\,\mathrm{cn}\left(u,k\right)\,$, $\,\mathrm{sn}\left(u,k\right)\,$,
 and $\,\mathrm{dn}\left(u,k\right)\,$ and use the derivative identities
\begin{align}
\frac{\mathrm{d}}{\mathrm{d}u}\left(\mathrm{cn}\left(u,k\right)\right) &= -\mathrm{sn}\left(u,k\right)\mathrm{dn}\left(u,k\right)~~,\\
\frac{\mathrm{d}}{\mathrm{d}u}\left(\mathrm{sn}\left(u,k\right)\right) &= \mathrm{cn}\left(u,k\right)\mathrm{dn}\left(u,k\right)~~,\\
\frac{\mathrm{d}}{\mathrm{d}u}\left(\mathrm{dn}\left(u,k\right)\right) &= -k^{2}\mathrm{cn}\left(u,k\right)\mathrm{sn}\left(u,k\right)~~,
\end{align}
 to obtain
 \begin{align}
 \mathrm{sn}\left(u,k\right)\mathrm{dn}\left(u,k\right) &=
 \frac{\pi^{2}}{kK\left(k\right)^{2}}\sum_{n=0}^{\infty}\frac{\left(2n+1\right)q^{n+\frac{1}{2}}}{1+q^{2n+1}}\sin\left(\frac{\left[2n+1\right]\pi u}{2K\left(k\right)}\right)~,
 \label{eq:mixedbegin}\\
 \mathrm{cn}\left(u,k\right)\mathrm{dn}\left(u,k\right) &=
 \frac{\pi^{2}}{kK\left(k\right)^{2}}\sum_{n=0}^{\infty}\frac{\left(2n+1\right)q^{n+\frac{1}{2}}}{1-q^{2n+1}}\cos\left(\frac{\left[2n+1\right]\pi u}{2K\left(k\right)}\right)~,
 \label{eq:mixedmid}\\
 \mathrm{sn}\left(u,k\right)\mathrm{cn}\left(u,k\right) &=
 \frac{2\pi^{2}}{k^{2}K\left(k\right)^{2}}\sum_{n=0}^{\infty}\frac{\left(n+1\right)q^{n+1}}{1+q^{2n+1}}\sin\left(\frac{\left[n+1\right]\pi u}{K\left(k\right)}\right)~,
 \label{eq:mixedend}
 \end{align}
 where $\,q\,$ is the Jacobi Nome defined as
 \begin{equation}
 q\,=\,\exp\left(-\frac{\pi K\left(\sqrt{1-k^{2}}\right)}{K\left(k\right)}\right)\,~,
 \end{equation}
 with $\,K\left(k\right)\,$ being the complete elliptical integral of the first kind
 \begin{equation}
 K\left(k\right)=\intop_{0}^{\pi/2}\left(1-k^{2}\sin\theta\right)^{-1/2}\mathrm{d}\theta\,~.\label{eq:q}
 \end{equation}

 Expansions for the Jacobi elliptic functions squared are similarly found. The first derivation of the Fourier development for $\,\mathrm{sn}^{2}\left(u,k\right)\,$ dates back to
 Jacobi in his treatise on elliptic functions which, by further appealing to Legendre's relation for complete elliptical integrals \citep{whittaker1996course}, is given by
 (see Section 41, p. 110 of \cite{jacobi1969fundamenta})
 \begin{equation}
 \mathrm{sn}^{2}\left(u,k\right)=\frac{1}{k^{2}}\left(1-\frac{E\left(k\right)}{K\left(k\right)}\right)-\frac{2\pi^{2}}{k^{2}K\left(k\right)^{2}}\sum_{n=1}^{\infty}\frac{nq^{n}}{1-q^{2m}}\cos\left(\frac{m\pi u}{K\left(k\right)}\right), \label{eq:sn2nome}
 \end{equation}
with $E\left(k\right)\,$ being the complete integral of the second kind
 \begin{equation}
 E\left(k\right)=\intop_{0}^{\pi/2}\left(1-k^{2}\sin\theta\right)^{1/2}\mathrm{d}\theta\,~.
 \end{equation}
 Expansions for $\mathrm{cn}^{2}\left(u,k\right)$ and $\mathrm{dn}^{2}\left(u,k\right)$ are obtained by using the fundamental elliptical identities
 \begin{align}
 \mathrm{cn}^{2}\left(u,k\right)&=1-\mathrm{sn}^{2}\left(u,k\right)\,~,\\
 \mathrm{dn}^{2}\left(u,k\right)&=1-k^{2}\mathrm{sn}^{2}\left(u,k\right)\,~.
 \end{align}

 Some comments regarding expansions (\ref{eq:mixedbegin} - \ref{eq:mixedend}) and \eqref{eq:sn2nome} are in order. First, we note that all expansions are bounded in the region $k\in\left[0,1\right)$, with $K\left(1\right)$ corresponding to
 positive imaginary
 infinity,
 thereby causing the expansions to break down. This regime corresponds to a transitioning of the Jacobi elliptic functions from doubly-periodic to non-periodic hyperbolic functions.
 These expansions only provide an apt analytic description of the rotator's kinematics provided we do not consider the separatrix dividing the LAM and SAM rotational behaviours.

 Second, we note that, given the defintion of the Nome as an exponential in \eqref{eq:q}, all the listed expansions exponentially converge to the desired Jacobi elliptic function. However, as we consider the $k\rightarrow 1$ regime where the expansions break down, one needs more terms in order to properly encompass the increasingly
 non-periodic nature of the elliptic functions.

 The third and most important point regards \eqref{eq:sn2nome}: This expansion is valid for all $u$. Despite $\mathrm{sn}^{2}\left(u,k\right)\in\left[0,1\right]$ and the terms
 of the Nome expansion being cosines capable of assuming negative values, the expansion remains within $\left[0,1\right]$ for $k\in\left[0,1\right)\,$. Indeed, the inequality
 \begin{equation}
 \frac{1}{k^{2}}\left(1-\frac{E\left(k\right)}{K\left(k\right)}\right)>\frac{2\pi^{2}q}{k^{2}K\left(k\right)^{2}\left(1-q\right)}~~,
 \end{equation}
 suggests that the first oscillatory term has an amplitude less than the constant zeroth-order term. Furthermore, this will not change with the inclusion of higher order terms,
 given the exponential decay of the series, so that the expansion must be within the region $\left[0,1\right]$ for $k\neq 1$. The important consequence is that the obtained viscoelastic stresses will remain valid for all time, rather than being piecewise functions.

 With the introduction of the Nome expansions, we are now able to Laplace transform the elastic stress by integrating term by term according to \eqref{eq:Laplacedef}.
 Given that we have solved the problem in the infinitesimal strain regime, the easiest means of computation involves transforming the entries of $\boldsymbol{S}^{\left(00\right)}$,
 since all other contributions of $\boldsymbol{\sigma}$ are related linearly to them. Therefore, as an example, consider implementing step 1 of our method to $S_{13}^{\left(00\right)}$
 given in \eqref{eq:example}: The only time-dependent term is $B_{13}$ defined in \eqref{eq:BB} and is therefore the only term which must be transformed. We integrate the expansion
 \eqref{eq:mixedmid}
 \begin{align}
 \mathcal{L}\left\{B_{13}\right\}&=
 ~C\mathcal{L}\left\{\mathrm{cn}\left(\omega^{\left(R\right)}t,k^{\left(R\right)}\right)\mathrm{dn}\left(\omega^{\left(R\right)}t,k^{\left(R\right)}\right)\right\}
 ~\\
 &=~C\sum_{n=1}^{\infty}\frac{\left(2n+1\right)q^{n+1/2}}{1-q^{2n+1}}\left(\frac{\textstyle s}{\textstyle \left[\frac{\textstyle\left(2n+1\right)\omega^{\left(R\right)}}{\textstyle
 2K\left(k\right)}\right]^{2}+s^{2}}\right)~~,
 \end{align}
 where we have set $\,t_{0}=0\,$, while the superscript $\,R\,$ is either $\,L\,$ or $\,S\,$, for LAM and SAM rotation respectively.
 The nome is calculated as $\,q=q\left(k^{\left(R\right)}\right)\,$, and the overall factor is
 \begin{equation}
 C~=~\frac{\mp~2~\Omega_{1}^{\left(0\right)}\,\Omega_{3}^{\left(0\right)}}{1~+~h_{1}^{2}\,h_{2}^{2}}\,~,
 \end{equation}
 with
 \begin{equation}
\Omega_{1}^{(0)}~=~\frac{\left|\boldsymbol{J}\right|}{I_{11}}~
         \sqrt{\frac{\left(\mathcal{B}^{\left(R\right)}\,I_{33}-I_{22}\right)I_{11}}{\left(I_{33}-I_{11}\right)I_{22}}} \,~, \; \; \Omega_{3}^{(0)}~=~\frac{\left|\boldsymbol{J}\right|}{I_{33}}~
         \sqrt{\frac{\left(I_{22}-\mathcal{B}^{\left(R\right)}\,I_{11}\right)I_{33}}{\left(I_{33}-I_{11}\right)I_{22}}}  \,~.
 \end{equation}

 The fictitious stresses in the complex frequency space are then found by supposing a linear viscoelastic rheology which defines the operators $P_{1}$, $U_{1}$, $P_{2}$, and $U_{2}$
 according to \eqref{eq:veDt} and \eqref{eq:veVt}. These operators are transformed into algebraic operators $\widehat{P}_{1}$, $\widehat{U}_{1}$, $\widehat{P}_{2}$, and
 $\widehat{U}_{2}$ by the Laplace transform property \eqref{eq:Laplaceprop}. Then, we substitute $\nu\rightarrow\widehat{\nu}_{VE}$ as defined in \eqref{eq:genPRs}. Explicitly, the
 corresponding fictitious stress of \eqref{eq:example} is given by
 \begin{align}
 \nonumber
 &\widehat{S}_{13}^{\left(00\right)}\left(\widehat{\nu}_{VE},s\right) =&\nonumber\\
 &\frac{\textstyle \rho \,h_{1}^{2}\,h_{2}^{2}\mathcal{L}\left\{B_{13}\right\}\left(h_{1}^{2}\left(h_{2}^{2}(\widehat{\nu}_{VE}+1)+2\right)+
 \left(2h_{2}^{2}+1\right)(\widehat{\nu}_{VE}+1)\right)}
 {\textstyle 2~h_{1}^{2}\,\left(h_{2}^{2}\,(\widehat{\nu}_{VE}+1)+2\right)+2\left(3h_{2}^{2}+1\right)(\widehat{\nu}_{VE}+1)}~~.\label{eq:examplefict}
 \end{align}

 \subsection{Determining viscoelastic stresses by Residue Theory}\label{sec:res}

 The last step of determining the viscoelastic stress involves taking the fictitious stress in Laplace space and inverse transforming it by the operator $\mathcal{L}^{-1}$.
 For linear viscoelastic rheologies, the transformed deviatoric operators $\widehat{P}_{1}$ and $\widehat{U}_{1}$ and the volumetric operators $\widehat{P}_{2}$ and $\widehat{U}_{2}$
 are integer polynomials of $s$. This aspect allows us to obtain a simplified form for the inverse transform (see Appendix \ref{sec:proof} for proof)
 \begin{equation}
 \mathcal{L}^{-1}\left\{\widehat{\alpha}\left(s\right)\right\}=\sum\mathrm{Res}\left[e^{st}\widehat{\alpha}\left(s\right)\right]\;, \label{eq:inverseLap}
 \end{equation}
 where the right hand side is the sum of all complex residues or contributions from the singularities of $e^{st}\widehat{\alpha}\left(s\right)$, and where the highest derivatives of
 the deviatoric and volumetric viscoelastic operators satisfy $m_{1}+n_{2} \leq m_{2}+n_{1}$, as defined in the viscoelastic problem \eqref{eq:Pop} and \eqref{eq:Uop}.

Given that our case exhibits integer powers of $s$, these singularities are poles. As a reminder, a function $f$ which is zero at $s_{0}$ becomes a pole for $1/f\left(s_{0}\right)$. Such poles are removable singularities, being multiplied by powers of $\left(s-s_{0}\right)$ until $\left(s-s_{0}\right)^{n}/f\left(s_{0}\right)$ produces a finite value for the lowest possible integer $n$. The residues of such $n$th order poles are computed by \citep{mitrinovic1984cauchy}
\begin{equation}
\mathrm{Res}\left[e^{st}\widehat{\alpha}\left(s\right)\right]=\frac{1}{\left(n-1\right)!}\lim_{s\rightarrow s_{0}}\frac{\mathrm{d}^{n}}{\mathrm{d}s^{n}}\left[\left(s-s_{0}\right)e^{st}\widehat{\alpha}\left(s\right)\right]\;. \label{eq:residue}
\end{equation}
The pole singularities of the fictitious stress have a very nice physical meaning: If the $s_{0}$ where the singularity occurs is purely real, this defines the exponential relaxation timescale of the viscoelastic stress as per \eqref{eq:residue}, whereas a purely imaginary $s_{0}$ defines the frequency at which the stress oscillates, given that it results in a complex exponential.

The method of computation to determine viscoelastic stress is as follows: Having found the fictitious stress $\widehat{\boldsymbol{\sigma}}\left(\widehat{\nu}_{VE},s\right)$, one must determine the values for the Laplace variable $s$ whereby pole singularities occur. These values can be determined by demanding the denominator of the stresses vanish, however, one should ensure that these solutions do result in pole singularities by taking the limit of the fictitious stress as it approaches the value for $\widehat{\nu}_{VE}$ or $s$.  In general, there are two sources of pole singularities: The first is from the Laplace transformed inertial forcing and self-gravitation $\mathcal{L}\left\{B_{ij}\right\}$ which, for our previous example \eqref{eq:examplefict}, occurs at
\begin{equation}
s_{0}=\pm\frac{\mathrm{i}\left(2n+1\right)\omega^{\left(R\right)}}{2K\left(k\right)}\;,\label{eq:firstcon}
\end{equation}
where $\mathrm{i}=\sqrt{-1}$, yielding the result that the poles are of order $2$ and are purely imaginary so that they define the frequency at which the stress oscillates.

The second contribution to the singularities can be most conveniently determined by demanding the denominator is zero and solving the resultant equation in terms of the transformed Poisson ratio $\widehat{\nu}_{VE}$ which, for our example, is
\begin{equation}
\widehat{\nu}_{VE}=-1-\frac{2h_{1}^{2}}{1+\left(3+h_{1}^{2}\right)h_{2}^{2}}\;.
\end{equation}

By supposing a linear rheology, one can then solve for the corresponding $s_{0}$ by using $\eqref{eq:genPRs}$. To illustrate, consider the rotator being described by a Kelvin-Voigt rheology under deviatoric deformations, so that
\begin{equation}
\widehat{P}_{1}=1\;, \qquad \widehat{U}_{1}=\mu+\eta s\;,
\end{equation}
where $\mu$ is the modulus of elasticity and $\eta$ is the viscosity, whilst it remains elastic under volumetric deformation so that $\widehat{P}_{2}=1$ and $\widehat{U}_{2}=3\mathcal{K}$, with $\mathcal{K}$ being the volumetric elastic modulus. In this case, the corresponding singularity occurs at
\begin{equation}
s_{0}=-\eta\left(\frac{6 K \left(\left(h_{1}^2+3\right) h_{2}^2+h_{1}^2+1\right)}{h_{1}^2 \left(h_{2}^2+4\right)+3 h_{2}^2+1}+\mu\right)\;, \label{eq:secondcon}
\end{equation}
which is a first order pole singularity that is purely real and therefore defines the exponential relaxation factor incorporating the material properties and geometry of the rotator.

The inverse Laplace transform of \eqref{eq:examplefict} is finally calculated by finding the residues of the pole singularities \eqref{eq:firstcon} and \eqref{eq:secondcon}, according 
to \eqref{eq:residue}. The benefit of working in Laplace space has reduced the problem of inverse transformation, and therefore finding the viscoleastic stresses in real time, to finding the solution of a polynomial and differentiating, rather than having to compute an integral which, in general, is more difficult to do analytically. This three step correspondence process of Laplace transforming, finding the fictitious stress in complex frequency space, and inverse transforming by determining the residues is done for each entry of $\boldsymbol{S}^{\left(00\right)}$ in order to obtain the total viscoelastic stress $\sigma_{VE}\left(t\right)$.

\section{Calculation of precession relaxation timescales}

 With the viscoelastic stresses found, we proceed with calculating the power dissipated due to internal stresses and therefore determine the timescale necessary to relax tumbling.
 For a continuum lacking internal heat sources, the dissipated power $\,P\left(t\right)\,$ comprises two parts: one owing to the work of the traction vector inside the body $\,\boldsymbol{t}\,$; another is due to the work of the body forces per unit mass over the volume $\,\boldsymbol{f}\,$
  \begin{equation}
 P(t)~=~ \int_S \boldsymbol{t}(\boldsymbol{x},t)
 \cdot \frac{{\partial\boldsymbol{u}}(\boldsymbol{x},t)}{\partial t}~dS + \int_V \rho~\boldsymbol{f}(\boldsymbol{x},t) \cdot \frac{{\partial\boldsymbol{u}}(\boldsymbol{x},t)}{\partial t}~dV,
 \label{51}
 \end{equation}
 with $\boldsymbol{u}$ being the displacement field. The coordinates $\boldsymbol{x}$ are those of Euler, so 
 $dV \equiv d^3 \boldsymbol{x}$ is an Eulerian (deformed) element of volume, while $\,dS\,$ is an element of the deformed surface.  
 
We use Cauchy's Law to impose a relationship between the traction vector and the stress, so that $\boldsymbol{t}=\boldsymbol{\sigma}\cdot\boldsymbol{n}$, and apply the Divergence Theorem to obtain
\begin{align}
P(t)&~=~\int_V \nabla\cdot\left(\boldsymbol{\sigma}
 \cdot \frac{{\partial\boldsymbol{u}}(\boldsymbol{x},t)}{\partial t}\right) + \rho~\boldsymbol{f}(\boldsymbol{x},t) \cdot \frac{{\partial\boldsymbol{u}}(\boldsymbol{x},t)}{\partial t}~dV,\\
 &~=~\int_V \boldsymbol{\sigma}:\nabla\left(\frac{{\partial\boldsymbol{u}}(\boldsymbol{x},t)}{\partial t}\right)+\left(\nabla\cdot\boldsymbol{\sigma}+\rho\boldsymbol{f}\left(\boldsymbol{x},t\right)\right)\cdot\frac{{\partial\boldsymbol{u}}(\boldsymbol{x},t)}{\partial t}~dV,
\end{align}
where $\,\left(:\right)\,$ is the double dot product between tensors.

Since $\,\nabla\cdot\boldsymbol{\sigma}\,+\,\rho\boldsymbol{f}\left(\boldsymbol{x},t\right)=0\,$ from the balance of linear momentum under the quasi-rigid and adiabatic approximations, the remaining term can be written down as
 \begin{equation}
 P(t)~=~\,\int_V\boldsymbol{\sigma}:\frac{\partial}{\partial t}\left(\nabla\boldsymbol{u}(\boldsymbol{x},t)\right)\,dV~~,
 \label{98}
 \end{equation}
which can be further simplified, upon using the symmetric property of the stress tensor, as 
 \begin{align}
\boldsymbol{\sigma}:\frac{\partial}{\partial t}\left(\nabla\boldsymbol{u}(\boldsymbol{x},t)\right)&=
 \frac{\textstyle 1}{\textstyle 2}\boldsymbol{\sigma}:\frac{\partial}{\partial t}\left(\nabla\boldsymbol{u}(\boldsymbol{x},t)\right)+
 \frac{\textstyle 1}{\textstyle 2}\boldsymbol{\sigma}^T:\frac{\partial}{\partial t}\left(\nabla\boldsymbol{u}(\boldsymbol{x},t)\right)^T,
 ~\nonumber\\
 &= \frac{1}{2}\boldsymbol{\sigma}:\frac{\partial}{\partial t}\left[\nabla\boldsymbol{u}\left(\boldsymbol{x},t\right)+\left(\nabla\boldsymbol{u}\left(\boldsymbol{x},t\right)\right)^{T}\right],\nonumber\\
 &= \boldsymbol{\sigma}:\frac{\partial\boldsymbol{\varepsilon}}{\partial t},
 \label{99}
 \end{align} 
 with the infinitesimal strain tensor defined as
 $$
 \boldsymbol{\varepsilon}\,\equiv\,\frac{\textstyle 1}{\textstyle 2}\left[\nabla\boldsymbol{u}+\left(\nabla\boldsymbol{u}\right)^{T}\right]~~.
 $$ 
 Inserting (\ref{99}) in (\ref{98}), we arrive at an expression for the power arising from internal stresses and strains
 \begin{align}
 P\left(t\right)&=~~\int_{V}\boldsymbol{\sigma}:\frac{\partial \boldsymbol{\varepsilon}}{\partial t}\mathrm{d}V,\nonumber\\
 &=~~\int_{V}\left(\boldsymbol{\sigma}_{D}+\boldsymbol{\sigma}_{V}\right):\frac{\partial}{\partial t}\left(\boldsymbol{\varepsilon}_{D}+\boldsymbol{\varepsilon}_{V}\right)\mathrm{d}V\,~,
 \label{eq:P}
 \end{align}
 where we have decomposed each tensor into their deviatoric and volumetric parts.

For analytic tractability in determining the NPA tumbling relaxation, we average the power dissipation over a precession period. Given the linear dependence of $\boldsymbol{\sigma}$ on the oscillatory inertial forces through $\mathbb{B}$ and the linear nature of the viscoelastic rheology, the double dot product in \eqref{eq:P} will feature quadratic terms in the inertial forcing. In light of the Nome expansions introduced in Section \ref{sec:Nome}, there will be couplings between different frequencies of forcing. It is not feasible to determine the period of all of these contributions, therefore, noting that all the expansions are multiples of the lowest fundamental frequency, we take the time average over the largest possible period of oscillation, because we are guaranteed that all couplings in the stress power will have completed at least one cycle. The oscillation with the largest period is derived from the Nome expansions for the mixed Jacobi elliptic functions \eqref{eq:mixedbegin} and \eqref{eq:mixedmid}, with $n=0$, yielding a period of $T=4K\left(k^{\left(R\right)}\right)/\omega^{\left(R\right)}$, according to the kinematics of the rotator determined in Section \ref{sec:solutions}, and $R=\left\{L,S\right\}$ for LAM and SAM rotation respectively.

Additional consideration must be given to  the exponential relaxation factors obtained from the singularities in the stress, such as the finding of \eqref{eq:secondcon} described in Section \ref{sec:res}, as these factors are non-periodic in time. To this end, we employ the previous assumption that the body has been rotating for a sufficiently long enough time to obtain a constant shape of revolution, which allows us to reduce the viscoelastic exponential relaxation to either zero or unity, depending on the linear rheology used and the material properties. For example, we note that \eqref{eq:secondcon}, which assumed a Kelvin-Voigt rheology, can be written as:
\begin{equation}
s_{0}=-\eta\left(\mathcal{YK}+\mu\right),
\end{equation}
for $\mathcal{Y}>0$ being a constant, which would result in the exponential relaxation factor $\exp\left(-\eta\left(\mathcal{YK}+\mu\right)t\right)$, according to \eqref{eq:inverseLap}. This factor would vanish for long times in the instance of $\eta\rightarrow \infty$ for elastic or very cold solids, or be unity for highly dissipating materials whereby $\eta\rightarrow 0^{+}$.

This approximation removes the non-periodic time contributions from the viscoelastic stresses and, coupled with our earlier comments on the Jacobi Nome expansions, enables us to write the time-average of the power $P_{\text{avg}}$ as
\begin{equation}
P_{\text{avg}}=\frac{1}{T}\int_{t_{0}}^{T+t_{0}}\int_{V}\left(\boldsymbol{\sigma}_{D}+\boldsymbol{\sigma}_{V}\right):\frac{\partial}{\partial t}\left(\boldsymbol{\varepsilon}_{D}+\boldsymbol{\varepsilon}_{V}\right)\mathrm{d}V\mathrm{d}t. \label{eq:Pavg}
\end{equation}

 Lastly, to calculate the decay of the wobbling angle $\,\theta_{max}^{\left(R\right)}\,$, we employ the relation \citep{breiter2012stress,frouard2017precession}
 \begin{equation}
 \frac{\mathrm{d}\theta_{max}^{\left(R\right)}}{\mathrm{d}t}\;=\;
 \frac{-P_{\text{avg}}}{\left|\boldsymbol{J}\right|^{2}\left(\frac{\textstyle 1}{\textstyle I_{22}}-\frac{\textstyle 1}{\textstyle I^{\left(R\right)}}\right)
 \;\sin\theta^{\left(R\right)}_{max}\;\cos\theta^{\left(R\right)}_{max}}\,~,
 \label{eq:wobblediss}
 \end{equation}
 with $\,I^{\left(L\right)}=I_{11}\,$ and $\,I^{\left(S\right)}=I_{33}\,$. Also be mindful that within the adiabatic approximation we separate timescales. In application to the above equation, this means that we assume both $\,\left|\boldsymbol{J}\right|\,$ and the moment of inertia matrix $\mathbb{I}$ are constant in time when integrating over the longest timescale, that of the relaxation.
  Then, noting that $\,P_{\text{avg}}\,$ depends directly on the wobbling angle $\,\theta_{max}^{\left(R\right)}\,$, we can rearrange the expression to find the time necessary
 for the rotator's maximal wobbling angle to decay from $\,\theta_{i}\,$ to $\,\theta_{f}\,$
 \begin{equation}
 t_{\text{relax}}^{\left(R\right)}\;=\;-\;\left|\boldsymbol{J}\right|^{2}\left(\frac{1}{I_{22}}-\frac{1}{I^{\left( R\right)}}\right)
 \int_{\theta_{i}}^{\theta_{f}}\frac{\sin\theta^{\left(R\right)}_{max}\;\cos\theta^{\left(R\right)}_{max}\;\mathrm{d}\theta^{\left(R\right)}_{max}}{P_{\text{avg}}}\,\;. \label{eq:trelax}
 \end{equation}

We comment on formulae \eqref{eq:Pavg} and \eqref{eq:wobblediss}: Given that we previously assumed that the triaxial geometry satisfied $I_{11}<I_{22}<I_{33}$, we note that in LAM rotation, \eqref{eq:wobblediss} predicts that the the wobbling angle of the rotator will increase, whilst in SAM rotation, it will decrease, coinciding with our description of the precession relaxation process in Section \ref{sec:preliminary}. This is true provided our assumptions regarding the body being heatless, isolated, and freely rotating hold, as this ensures that $P_{\text{avg}}\geq 0$.

 Regarding \eqref{eq:Pavg}, we remark that the employment of time-averaging results imposes a limitation on analytical modeling of the rotational
 behaviour close to the transition between LAM and SAM modes. Time-averaging requires that the function be periodic in time,
 while at the separatrix the maximal wobble angle is $\,\theta_{max}^{\left(R\right)}=90^{\circ}\,$. This value of the angle corresponds to
 $\,k^{\left(R\right)}=1\,$ which renders
 \begin{equation}
 \mathrm{cn}\left(u,1\right)=\mathrm{sech}\left(u\right), \qquad \mathrm{sn}\left(u,1\right)=\mathrm{tanh}\left(u\right), \qquad \mathrm{dn}\left(u,1\right)=\mathrm{sech}\left(u\right).
 \end{equation}
 The previously doubly-periodic functions now become non-periodic at the separatrix, thus breaking the validity of time-averaging. This provides a
 mathematical reason for why we are only able to consider the precession relaxation purely within the LAM or the SAM regime, but not the transition
 between these two rotational behaviours.

 \section{`Oumuamua and Toutatis as ellipsoidal Maxwell bodies \label{sec:Oumuamua}}

We apply our previously developed theory to two examples, the interstellar 1I/2017
(`Oumuamua) and the planet-orbit crossing 4179 Toutatis, which represent the extremes of energy dissipation, with the former asteroid being almost non-dissipative,
and the latter being highly dissipative. We generate timescale estimates for the precession relaxation, based on experimentally observed data and, at the same time, vary the parameters which are unknown to develop an intuition for the role that the asteroid's geometry, mechanics, and material properties play in this process.

 We take the deviatoric deformation of the body to obey the Maxwell rheology, whilst the volumetric deformation is taken to be elastic. The former is taken to have the Maxwell rheology because it has been experimentally shown to be in good agreement with data, particularly for low temperature bodies such as `Oumuamua and Toutatis, whilst the latter is supposed because the bulk viscosity is many orders of magnitude larger than the shear viscosity for cold bodies. Therefore, for our case, the constitutive relations are
 \begin{align}
 \boldsymbol{\sigma}_{D}+\frac{\eta}{\mu}\frac{\partial\boldsymbol{\sigma}_{D}}{\partial t}&=\eta\frac{\partial\boldsymbol{\varepsilon}_{D}}{\partial t}\;,\label{eq:appD}\\
 \boldsymbol{\sigma}_{V} &= 3\mathcal{K}\boldsymbol{\varepsilon}_{V}\;,\label{eq:appV}
 \end{align}
 where $\eta$ is the shear viscosity, $\mu$ is the shear elasticity modulus, and $\mathcal{K}$ is the bulk elastic modulus.

 We further reduce the number of material parameters in our system by setting the Poisson ratio $\nu=1/4$, a value most often assumed by cold solids, as argued by
 \cite{efroimsky2000inelastic}. For a viscoelastic material, a constant Poisson ratio is meaningless, given the dynamic nature of the constitutive relations \eqref{eq:appD} and \eqref{eq:appV}. Nevertheless, we can still impose this constraint by taking the elastic limit $\eta\rightarrow\infty$ of the generalised Poisson ratio \eqref{eq:genPR}, which renders:
 \begin{equation}
 \nu\sim\frac{3\mathcal{K}-\mu}{6\mathcal{K}-\mu}\,~.
 \end{equation}
 Then the assumption $\nu=1/4$ will give us a relation between the bulk and shear moduli as $\,\mathcal{K}=5\mu/6\,$. This is a tolerable approximation, because for realistic materials the
 values of $\,\mathcal{K}\,$ and $\,\mu\,$ are not radically different, in contrast from the values of the bulk and shear viscosities which differ from one another greatly.

 For the case of a Maxwell rheology with this reduction of parameters, the nonperiodic exponential viscoelastic relaxation terms derived from the stress solution can be written as
 $\exp\left(-\mathcal{Z}\mu t/\eta\right)$, where $\mathcal{Z}\in\left(0,1\right)$. Therefore, determining whether this factor contributes to the time-averaged power, according to \eqref{eq:Pavg}, is a matter of comparing the viscosity, which is inversely proportional to the energy dissipation rate, with the mechanical resistance to deformation.

 Lastly, to obtain the forthcoming results, we employ the first $5$ terms in the Nome expansions discussed in Section \ref{sec:Nome}, and analytically determine the viscoelastic
 stresses of the rotator by following our derived theory, with assistance from a symbolic algebra package. From this, determination of the precession relaxation time by means of
 \eqref{eq:trelax} is obtained through standard Riemann numerical integration.

\subsection{Relaxation of `Oumuamua}

 We now proceed with calculating the time estimates for `Oumuamua. Previous work by \cite{fraser2018tumbling} found that, by using the precession decay estimate provided by
 \cite{burns1973asteroid}, the characteristic time for the interstellar asteroid to reach its minimal energy rotation state was approximately $10^{10}-10^{12}\;\text{years}$,
 with the lower bound supposing an icy material composition and the upper bound corresponding to a composition of rock typical of a C-type asteroid. This result provides a
 starting point for further investigation with the more general theory presented here.

 \begin{table}
 \begin{center}
 \begin{tabular}{|c|c|}
 \hline
 Parameter & Numerical Value\tabularnewline
 \hline
 \hline
 Newton's gravitational constant & $G=6.674\times10^{-11}\;\text{m}^{3}\text{kg}^{-1}\text{s}^{-2}$\tabularnewline
 \hline
 Largest semi-major axis length & $a=115\;\text{m}$\tabularnewline
 \hline
 First aspect ratio & $h_{1}=\frac{\textstyle b}{\textstyle a}=\frac{\textstyle 15}{\textstyle 115}$\tabularnewline
 \hline
 Angular momentum  & $\left|\boldsymbol{J}\right|=5\times10^{7}\;\text{kg}.\text{m}^{2}\text{s}^{-1}$\tabularnewline
 \hline
 Shear modulus & $\mu=5\times10^{10}\;\text{Pa}$\tabularnewline
 \hline
 \end{tabular}
 \end{center}
 \caption{Experimentally reported or accepted estimates of parameters used to calculate the precession relaxation of 1I/2017 (\textquoteleft Oumuamua).}\label{tab:Oumuamua}
 \end{table}

 We use the parameters given in Table \ref{tab:Oumuamua}, where the dimensions and aspect ratio of the asteroid are taken from \cite{jewitt2017interstellar}. We employ a commonly used
 estimate of $\,\mu\,$ for monolith rocks \citep{rb} and calculate the estimate for the angular momentum by averaging the rotation around
 the longest axis and shortest axis, with mass densities varying between ice and rock of a C-type asteroid $1\times 10^{3}-2\times 10^{3}\,\text{kg}\, \text{m}^{-3}$ and the angular velocity reported
by \cite{bolin2017apo}. Furthermore, since \textquoteleft Oumuamua is a cold body, we expect its viscosity to exceed its shear modulus by orders of magnitude, whence the exponential relaxation term reduces to $\exp\left(-z\mu t/\eta\right)\sim 1$.

\begin{figure}
\begin{center}
\includegraphics[width=0.75\textwidth]{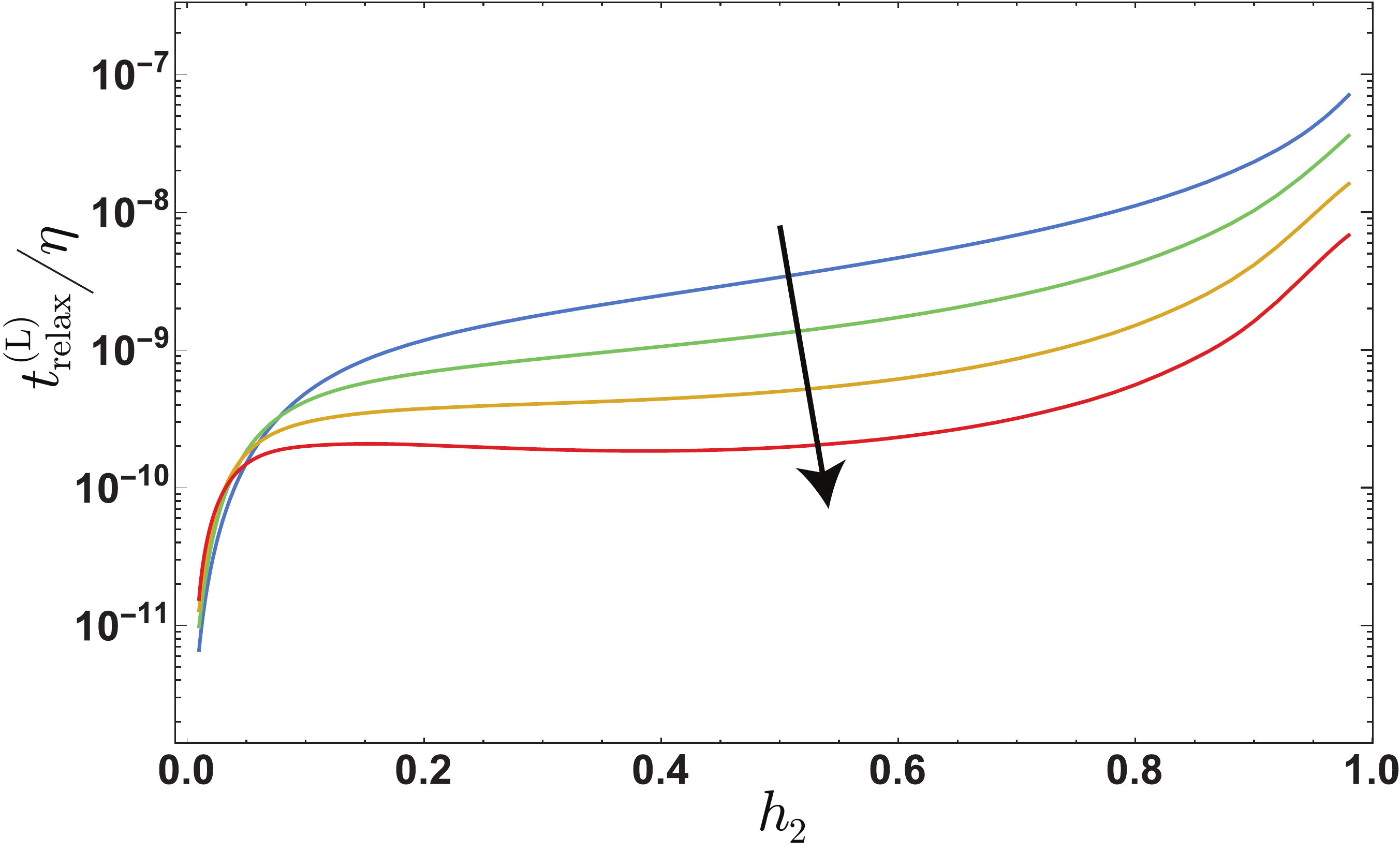}\\
\includegraphics[width=0.75\textwidth]{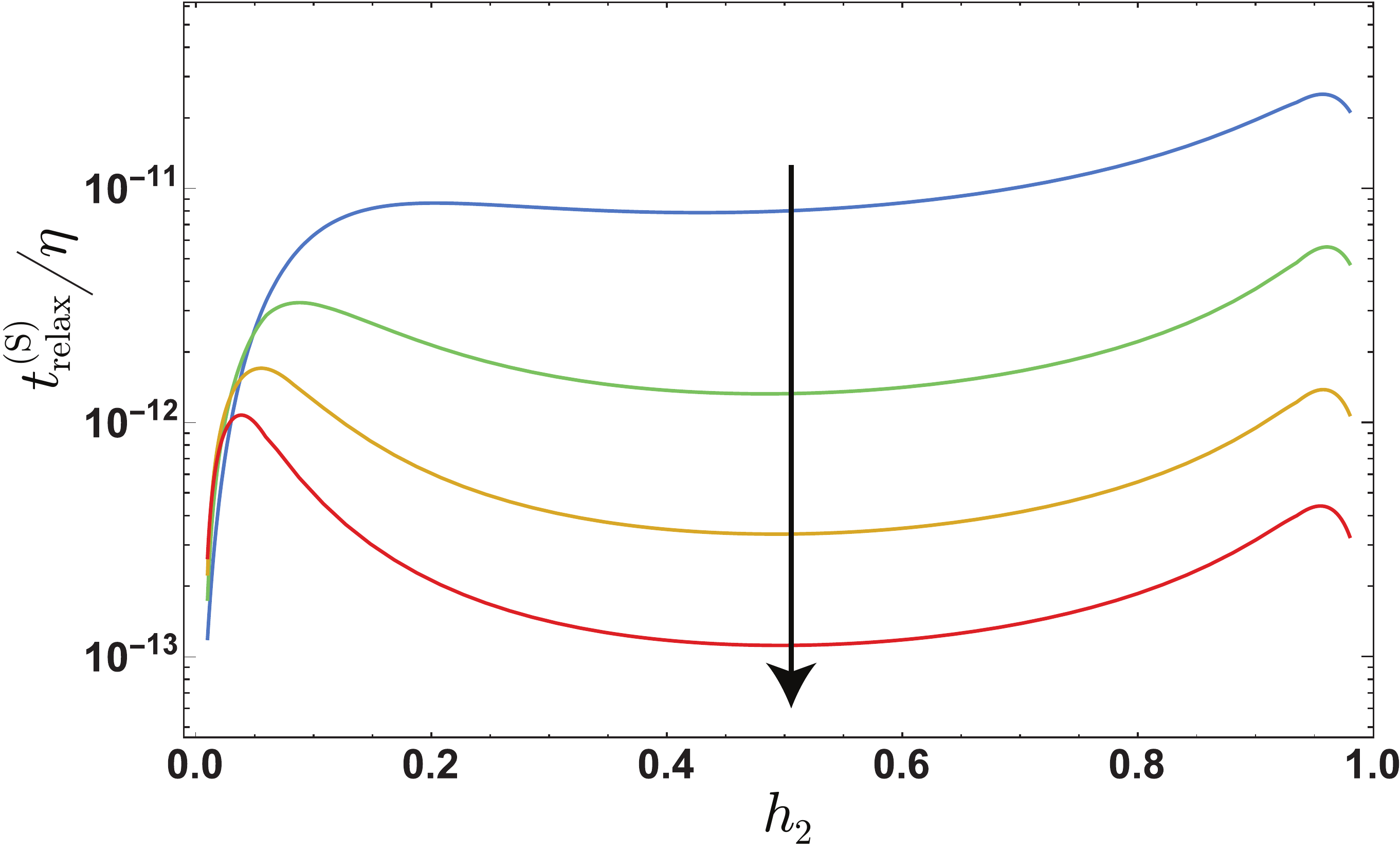}
\end{center}
\caption{Time rescaled with respect to material viscosity $t_{\text{relax}}^{\left(R\right)}/\eta$, in units $\text{yrs}.\text{Pa}^{-1}.\text{s}^{-1}$, necessary to relax \textquoteleft Oumuamua's maximum wobbling angle from $5^{\circ}$ to $85^{\circ}$ in LAM rotation (top) and $85^{\circ}$ to $5^{\circ}$ in SAM rotation (bottom) for varying second aspect ratio $h_{2}$ and mass densities $\rho=\left\{1\times 10^{3},1.5\times 10^{3},2\times 10^{3},2.5\times 10^{3}\right\}$ in units of kilograms per cubic meter. Arrows point in the direction of increasing mass density \label{fig:densityOumuamua}}
\end{figure}

Given that the mass density $\rho$ and second aspect ratio $h_{2}$ are not precisely known for \textquoteleft Oumuamua, we generate figures of the rescaled relaxation time $t_{\text{relax}}^{\left(R\right)}/\eta$, as the viscosity can be scaled out of the expression for $P_{\text{avg}}$ under the Maxwell rheology, necessary for the maximum wobbling angle $\theta_{\text{max}}^{\left(R\right)}$ to dissipate from $5^{\circ}$ to $85^{\circ}$ in LAM rotation and from $85^{\circ}$ to $5^{\circ}$ in SAM rotation with respect to $h_{2}$ for various values of $\rho$. In particular, we plot the relaxation time with respect to $h_{2}$ for mass densities $\rho=\left\{1\times 10^{3},1.5\times 10^{3},2\times 10^{3},2.5\times 10^{3}\right\}$, with units in kilograms per cubic meter, in Fig. \ref{fig:densityOumuamua}.

We comment on some important aspects of our results. First, noting that the total relaxation time from the maximum energy rotation state to the minimum was given by the sum of $t_{\text{relax}}^{\left(L\right)}+t_{\text{relax}}^{\left(S\right)}$, as discussed in Section \ref{sec:precession}, we observe that the dominant contribution occurs when \textquoteleft Oumuamua is in LAM rotation and is a factor of $10^2-10^3$ longer than its SAM counterpart. We reason that this aspect could be a result of the additional factor of $k$ describing the angular velocity around the $\boldsymbol{e}_{2}$ axis in LAM rotation, given in \eqref{eq:addkfactor}. This additional factor would result in the appearance of terms of $k^{2}$ in the power calculation, given that we are concerned with linear rheologies, which gives that the LAM time should be a factor of $1/k^{2}$ longer than the SAM time. For $k\in\left(0,1\right)$, this gives an approximate increase of $10^{2}$ over the integration region $\theta\in\left[5^{\circ},85^{\circ}\right]$.

The longest prediction for the relaxation time is approximately $10^{-7}\times\eta\;\text{yrs}$ for an almost oblate \textquoteleft Oumuamua, so the estimate found by \cite{fraser2018tumbling} is equivalent to setting $\eta\gtrapprox 10^{17}-10^{19}\;\text{Pa}.\text{s}$ in our theory, which physically corresponds to terrestrial planets where the mantle is hot and bound by pressure. Realistically, \textquoteleft Oumuamua is a very cold body which has a viscosity in the range $\eta\approx 10^{34}-10^{200}\;\text{Pa}.\text{s}$ or perhaps even higher, hence yielding a relaxation time-scale whose upper bound is $10^{23}-10^{193}\;\text{yrs}$; a substantially larger estimate than what the empirical $Q$-factor approach can predict. Given that the relaxation time-scale is significantly larger than the age of the universe, we can say with a high degree of certainty that, in the absence of an unlikely interstellar collision, `Oumuamua's NPA behavior will remain unchanged from its original formation.

Second, we remark on the effects of increasing \textquoteleft Oumuamua's mass density from ice to much denser rock in Fig. \ref{fig:densityOumuamua}. For both LAM and SAM rotation, the time necessary to dampen the rotational behaviour decreases with increasing mass density. The reason for this lies in the balance of linear momentum used to solve for the elastic stresses of the rotator \eqref{eq:BLM}; namely, that with increased mass density, the corresponding stresses must similarly increase, which then leads to greater power dissipation according to \eqref{eq:Pavg}.

Studying the dependence of the relaxation times with respect to the second aspect ratio $h_{2}$, we note that, for low densities, \textquoteleft Oumuamua in LAM rotation monotonically increases as the body becomes further oblate, whilst in SAM rotation, both local maximum values and minimum values of the relaxation time for particular values of $h_{2}$ exist. As the mass density increases to dense rock, the LAM relaxation time-scale curve loses its monotonicity and features a local minimum at $h_{2}\approx 0.3855$ for $\rho=2.5\times 10^{3}\;\text{kg}.\text{m}^{3}$ while the local maxima and minima in the low-density SAM curve become more pronounced, exhibiting a relaxation time maximum for low and high $h_{2}$ and a minimum for intermediate values of this second aspect ratio.

\subsection{Relaxation of Toutatis}

 Having studied a virtually nondissipative body, we now focus our attention on a rotating object which is highly dissipative, Toutatis. Observations on the structure of the asteroid
 made by the Chang'e-2 probe suggested that it is likely a rubble pile comprised of a number of loosely bound rocks under the effects of gravity \citep{huang2013ginger}. To obtain a
 reasonable estimate for its viscosity, we recall that analysis of the collapse of impact craters produces a value of about $\,\eta\,\approx\,2.4\times 10^8
 \;\text{Pa}.\text{s}\,$ for rubbles and near-rubbles \citep{melosh}.

 We use the values of the parameters as in Table \ref{tab:Toutatis}. Toutatis's dimensions are averages of those reported in \cite{huang2013ginger} and the
 NASA Jet Propulsion Laboratory's Database (https://ssd.jpl.nasa.gov/sbdb.cgi?sstr=2004179), whilst the values for the mass density were referenced from \cite{scheeres1998dynamics}.
 The angular momentum was derived by taking the rotational period to be $\,176\,$ hours per revolution \citep{warner} and averaging the angular momentum of an ellipsoid rotating
 around its largest and shortest axes.

\begin{table}
\begin{center}
\begin{tabular}{|c|c|}
\hline
Parameter & Numerical Value\tabularnewline
\hline
\hline
Largest semi-major axis length & $a=4.505\times 10^{3}\;\text{m}$\tabularnewline
\hline
First aspect ratio & $h_{1}=\frac{\textstyle b}{\textstyle a}=0.4909$\tabularnewline
\hline
First second ratio & $h_{2}=\frac{\textstyle c}{\textstyle b}=0.8250$\tabularnewline
\hline
Mass density & $\rho=2.1\times 10^{3}\;\text{kg}.\text{m}^{-3}$\tabularnewline
\hline
Angular momentum  & $\left|\boldsymbol{J}\right|=5.296\times10^{15}\;\text{kg}.\text{m}^{2}\text{s}^{-1}$\tabularnewline
\hline
Shear modulus & $\mu=5\times10^{10}\;\text{Pa}$\tabularnewline
\hline
Shear viscosity & $\eta=2.4\times 10^{8}\;\text{Pa}.\text{s}$\tabularnewline
\hline
\end{tabular}
\caption{Experimentally reported or accepted estimates of parameters used to calculate the precession relaxation of 4179 Toutatis.}\label{tab:Toutatis}
\end{center}
\end{table}

 The viscosity multiplied by a typical frequency $\chi$ (approximately $\,10^{-5}$ Hz) is many orders less than the rigidity, wherefore the characteristic viscoelastic relaxation exponential
 $\,\exp\left(-z\mu t/\eta\right)\,$ approaches zero very quickly. Combined with our assumption that Toutatis has been rotating for a sufficiently long time, it is reasonable to assume that the non-periodic exponential factor can be removed from the viscoelastic stress for the purpose of time-averaging.
 
Calculating the relaxation time for the maximum wobbling angle to dissipate from $5^{\circ}$ to $85^{\circ}$ in LAM and from $85^{\circ}$ to $5^{\circ}$ in SAM rotation, we obtain the values $4.3\times 10^{-8}\;\text{yr}$ and $2.4\times10^{-9}\;\text{yr}$ respectively. As with `Oumuamua, the relaxation time was longer in LAM than in SAM, however, the most noteworthy aspect of this result arises upon comparison with estimates presented in previous works \cite{burns1973asteroid} and \cite{efroimsky2000inelastic}: Namely, our estimate is several orders of magnitude smaller than these calculations, which lie in the range of $10^{9}-10^{11}\;\text{yr}$.

Our result is a direct consequence of the full mechanical treatment of the precession relaxation, not only by means of writing the power dissipation in less phenomenological parameters, such as the viscosity, but also including the effects of self-gravitation, known as pre-stressing. To illustrate where these differences arise, consider the relationship between the quality factor and the parameters of the Maxwell rheology (see Appendix D of \cite{frouard2017precession})
\begin{equation}
\mu Q = \eta\chi,
\end{equation}
where $\chi$ is the oscillating frequency of the stress defined, according to the angular frequencies of the Nome expansions \eqref{eq:sn2nome}, as
\begin{equation}
\chi \geqslant \frac{\pi\omega^{\left(R\right)}}{K\left(k^{\left(R\right)}\right)}\approx 10^{-5}\;\text{Hz},
\end{equation}
supposing the maximum value from LAM rotation and the fundamental base frequency, because this is the rotational mode which dominates the precession relaxation.

Consequently, using $\eta=2.4\times 10^{8}\;\text{Pa.s}$ from Table \ref{tab:Toutatis} yields that $\mu Q$ is of the order of $10^{3}\;\text{Pa}$ which is many orders of magnitude smaller than $10^{12}\;\text{Pa}-10^{13}\;\text{Pa}$ supposed by \cite{efroimsky2000inelastic} and \cite{burns1973asteroid} respectively. The corresponding relaxation time scale using our parameters from Table \ref{tab:Toutatis} into \eqref{1} ($\Omega = \chi \approx 10^{-5}\;\text{Hz}$) produces the estimate $\tau\approx 1\;\text{yr}$, which is still larger than the result derived from the full mechanical treatment. 

This final discrepancy arises as a result of neglecting pre-stress, which estimates such as \eqref{1} assume. Note however that this assumption breaks down for bodies which are highly energy dissipating (i.e. $\eta\chi \ll \mu$). To verify this, we repeat our full numerical calculation of the precession relaxation time, with $G=0$ to switch off the effects of self-gravity, and find the results $0.49\;\text{yr}$ and $0.080\;\text{yr}$ for LAM and SAM rotations respectively, the former of which corresponds to the previous estimate predicted by \eqref{1}. 

This example highlights the importance for a full viscoelastic theory to describe highly deformable and energy-dissipating bodies, as modeling these instances with the $Q$-factor approach will produce significant errors, especially when coupled with the omission of pre-stress.

\section{Reduction to an oblate geometry}

To better facilitate comparison with previous work and to derive a relatively simple expression for the characteristic relaxation time, we consider the limit where the geometry of the body is oblate. In this case, the principal semi-major axes of the rotator are $a = b \geq c$, so that there is only a single aspect ratio $h_{2}=h=\frac{c}{a}=\frac{c}{b}$, as $h_{1}=1$. As a result, the kinematic motion of the rotator is reduced to a single mode of tumbling that is parametrized by the wobbling angle $\theta\in\left[0^{\circ},90^{\circ}\right]$, with the upper and lower boundaries corresponding to the maximum and minimum energy rotational states respectively.

We repeat the steps to calculate the power dissipation and relaxation times as specified in Sections 2-5. We determine the forcing by solving the Euler equations \eqref{eq:rigid} to find the angular velocity components
\begin{align}
\Omega_{1} &= \frac{2\left|\boldsymbol{J}\right|}{I_{33}\left(1+h^{2}\right)}\sin\theta\cos\left(\omega\left(t-t_{0}\right)\right)\;\;,\\
\Omega_{2} &= \frac{2\left|\boldsymbol{J}\right|}{I_{33}\left(1+h^{2}\right)}\sin\theta\sin\left(\omega\left(t-t_{0}\right)\right)\;\;,\\
\Omega_{3} &= \frac{\left|\boldsymbol{J}\right|}{I_{33}}\cos\theta\;\;,
\end{align}
where $I_{33}=8\pi\rho ha^{5}/15$ in the oblate regime and $t_{0}$ is an integration constant, and determine the self-gravitation contributions by exactly integrating (\ref{eq:selfgravstart} - \ref{eq:selfgravend}):
\begin{align}
\gamma_{1}=\gamma_{2} & =\frac{G\pi h\rho\left(\pi-2h\sqrt{1-h^{2}}-2\sin^{-1}h\right)}{\left(1-h^{2}\right)^{3/2}}\;\;,\\
\gamma_{3} & =\frac{4\pi G\rho\left(\sqrt{1-h^{2}}-h\cos^{-1}h\right)}{\left(1-h^{2}\right)^{3/2}}\;\;.
\end{align}

Combining these results to construct the forcing matrix $\mathbb{B}$, as defined in \eqref{eq:BB}, and the elastic stresses in Appendix A for $h_{1}=1$, we calculate the viscoelastic stresses, assuming the same deviatoric Maxwell rheology and elastic volumetric material properties as in Section 6, for both the non-dissipative, $\eta\chi\gg \mu$, and highly dissipative, $\eta\chi\ll\mu$, limits to remove the non-periodic exponential terms from the stresses.

\begin{figure}
 \begin{center} 
\includegraphics[width=0.75\textwidth]{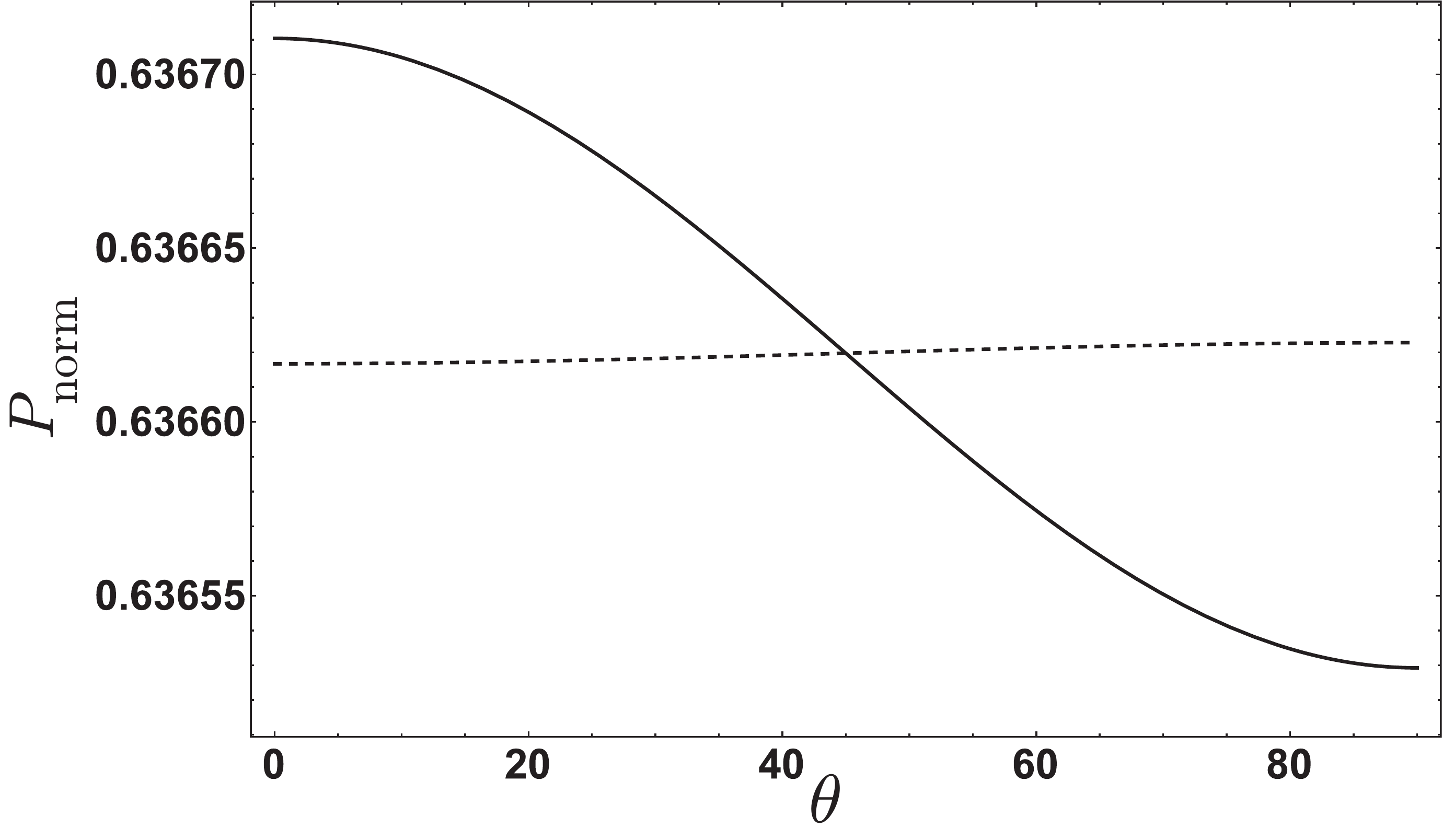}\\
\includegraphics[width=0.75\textwidth]{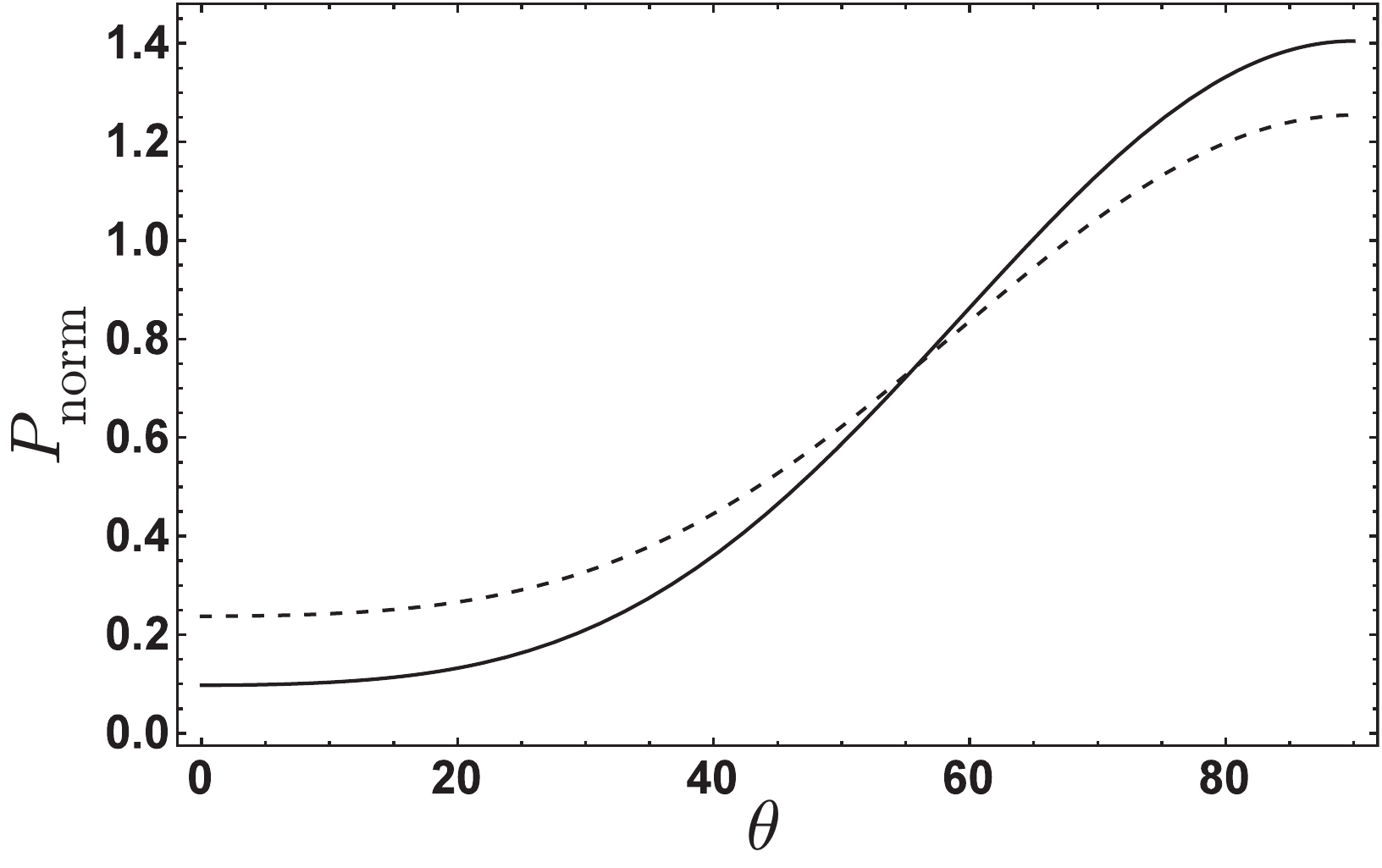}
 \end{center}
 \caption{Normalized power dissipation $P_{\text{norm}}$, in units $\text{J.s}^{-1}$, with respect to the wobbling angle $\theta$, in degrees, of an oblate `Oumuamua (solid line) and Toutatis (dashed line) assuming a deviatoric Maxwell rheology and volumetric elastic response, under the effects of self-gravitation (top) and without (bottom). In particular, the parameters $h=35/230$, $\rho=2.0\times 10^{3}\;\text{kg}.\text{m}^{-3}$, $\eta=10^{30}\;\text{Pa.s}$, and those presented in Table \ref{tab:Oumuamua} were used for Oumuamua, whilst $h=49/100$ and Table \ref{tab:Toutatis} were used for Toutatis.   \label{fig:PowerNorm}}
 \end{figure}

We determine the time averaged power according to \eqref{eq:Pavg} and plot the normalized power $P_{\text{norm}}$, whereby the curve is rescaled so that the enclosed area underneath is unity, with respect to the wobbling angle $\theta$ for an oblate `Oumuamua (with parameters $h=15/115$, $\rho=2.0\times 10^{3}\;\text{kg}.\text{m}^{-3}$, $\eta=10^{30}\;\text{Pa.s}$, and those given in Table \ref{tab:Oumuamua}) and Toutatis (with $h=49/100$ and parameters given in Table \ref{tab:Toutatis}), in Fig. \ref{fig:PowerNorm}. We also plot the the power dissipation both under the effects of self-gravitation and without to compare to previous results.

We discuss some important characteristics of our results: First, we observe that, under the effect of self-gravitation, the power dissipation for `Oumuamua appears reverse compared to \cite{breiter2012stress} and \cite{frouard2017precession}; namely, exhibiting a higher power dissipation for small wobbling angles, yet relatively low power dissipation for angles approaching $90^{\circ}$. On the contrary, Toutatis shows the usual dependence of power dissipation, however, the curve is relatively flat, almost independent of $\theta$, as a consequence of its high energy dissipation and the effects of pre-stress. When self-gravitation is switched off, we obtain curves more in line with figures presented in \cite{frouard2017precession} for both the non-dissipative and highly dissipative regimes, with the highest energy dissipation occurring when $\theta=90^{\circ}$ and monotonically decreasing, yet never to zero.

This observation leads to the second and important departure from the results of \cite{breiter2012stress} and \cite{frouard2017precession}. We argue that this result is reasonable for a Maxwell material as follows: For a body which is undergoing minimal energy rotation around its shortest principal axis, there is only a single non-zero and time independent angular velocity, which is $\Omega_{3}=\left|\boldsymbol{J}\right|/I_{33}$. In this case, the inertial effects of rotation are constant with respect to time, leading to $\mathbb{B}$ defined in \eqref{eq:BB} being time-independent, and subsequently the elastic stresses. By transformation and assuming the long time regime for a material, the viscoelastic stresses are similarly time independent. However, by the constitutive relation for a deviatoric Maxwell rheology \eqref{eq:appD}, a non-zero stress that is constant with respect to time will produce a non-zero strain rate, implying that the power dissipation \eqref{eq:Pavg} will not vanish for $\theta=0$.

We revisit comments made in Section \ref{sec:precession} regarding the validity of the work presented to only model the process of precession relaxation. After the minimal energy rotational state is obtained, power is still dissipated, which would theoretically manifest as a deformation in the geometry of the body. A constant inertial force would elongate the principal axes of the rotator, according to the Maxwell rheology, and would hence slow the angular velocity as a consequence of the conservation of angular momentum. We emphasize that an equilibrium geometry and hence constant angular velocity would eventually be obtained as a result of the inertial forcing, which stretches the body, balancing with self-gravitation, which compresses it, however, this is out of the scope of the present work and the assumptions regarding relaxation made in Section \ref{sec:preliminary}.

\subsection{Estimate for dampening timescale \label{sec:damp}}

Though we have derived a general theory for the tumbling relaxation of a rotator, we seek to derive relatively simple estimates for the characteristic dampening time of an oblate geometry, in a similar vein to the classic result by \cite{burns1973asteroid} given in \eqref{1}, yet including our full mechanical treatment of the problem. We strongly emphasize that the estimates derived here are only valid for bodies whose Poisson ratio $\nu$ is approximately $0.2$; namely, cold, non-rubble rotators.

We proceed by observing that the power dissipation curves under self-gravitation in Fig. \ref{fig:PowerNorm} are close to symmetric, regardless of whether higher energy loss corresponds to large or small wobbling angles. Given that we are interested in finding the dampening time necessary for $\theta$ to reduce from $90^{\circ}$ to $0^{\circ}$, this observation allows us to average the power over the wobbling angle without incurring a significant error, producing a $\theta$ independent function $\Psi$ that approximates $P_{\text{avg}}\left(\theta\right)$ and corresponds exactly for $P_{\text{avg}}\left(\theta\approx 45^{\circ}\right)$.

However, $\Psi$ is still algebraically complicated in our formalism, so we obtain a family of simplifying expressions by expanding the time and $\theta$ averaged power as a series in $\left(h-c\right)$ under the asymptotic limits $\eta \gg 1$ and $h\sim c$ for $c\in\left\{0.1,0.2,...,0.8,0.9\right\}$. This general expansion applies for both the non-dissipative and highly dissipative regimes and is given by
\begin{align}
\Psi & =\frac{1}{\eta}\left[\frac{X_{1}\left|\boldsymbol{J}\right|^{4}}{a^{13}\rho^{2}}+\frac{X_{2}G\left|\boldsymbol{J}\right|^{2}\rho}{a^{3}}+X_{3}a^{7}G^{2}\rho^{4}+\right.\nonumber\\
 & +\left(\frac{Y_{1}\left|\boldsymbol{J}\right|^{4}}{a^{13}\rho^{2}}+\frac{Y_{2}G\left|\boldsymbol{J}\right|^{2}\rho}{a^{3}}+Y_{3}a^{7}G^{2}\rho^{4}\right)\left(h-c\right)\nonumber\\
 & \left.+\left(\frac{Z_{1}\left|\boldsymbol{J}\right|^{4}}{a^{13}\rho^{2}}+\frac{Z_{2}G\left|\boldsymbol{J}\right|^{2}\rho}{a^{3}}+Z_{3}a^{7}G^{2}\rho^{4}\right)\left(h-c\right)^{2}\right]\nonumber\\
 &+O\left(\frac{1}{\eta^{2}},\left(h-c\right)^{3}\right)\;\;,\label{eq:Psi}
\end{align}
where $X$, $Y$, and $Z$ are series coefficients which we provide in Tables \ref{tab:seriesnondiss} and \ref{tab:seriesdiss} for various values of $c$.

\begin{table}
\begin{center}
\begin{tabular}{|c|c|c|c|c|c|}
\hline 
 & $c=0.1$ & $0.2$ & $0.3$ & $0.4$ & $0.5$\tabularnewline
\hline 
\hline 
$X_{1}$ & $84.0946$ & $9.34075$ & $2.31503$ & $0.788575$ & $0.32401$\tabularnewline
\hline 
$X_{2}$ & $-1.04476$ & $-0.770745$ & $-0.549492$ & $-0.379949$ & $-0.256657$\tabularnewline
\hline 
$X_{3}$ & $0.007732$ & $0.038778$ & $0.083662$ & $0.129058$ & $0.167433$\tabularnewline
\hline 
$Y_{1}$ & $-2591.43$ & $-154.300$ & $-27.8029$ & $-7.66355$ & $-2.63691$\tabularnewline
\hline 
$Y_{2}$ & $2.99468$ & $2.47885$ & $1.94809$ & $1.45217$ & $1.02763$\tabularnewline
\hline 
$Y_{3}$ & $0.195009$ & $0.404570$ & $0.469743$ & $0.425599$ & $0.340646$\tabularnewline
\hline 
$Z_{1}$ & $52215.4$ & $1596.14$ & $201.403$ & $44.4017$ & $13.0037$\tabularnewline
\hline 
$Z_{2}$ & $-2.45675$ & $-2.66034$ & $-2.60441$ & $-2.32400$ & $-1.90652$\tabularnewline
\hline 
$Z_{3}$ & $1.30830$ & $0.693772$ & $-0.002247$ & $-0.379551$ & $-0.418326$\tabularnewline
\hline 
\end{tabular}

\begin{tabular}{|c|c|c|c|c|}
\hline 
 & $c=0.6$ & $0.7$ & $0.8$ & $0.9$\tabularnewline
\hline 
\hline 
$X_{1}$ & $0.153401$ & $0.081651$ & $0.047722$ & $0.029891$\tabularnewline
\hline 
$X_{2}$ & $-0.171424$ & $-0.115133$ & $-0.079239$ & $-0.056703$\tabularnewline
\hline 
$X_{3}$ & $0.197889$ & $0.224858$ & $0.256348$ & $0.302293$\tabularnewline
\hline 
$Y_{1}$ & $-1.05142$ & $-0.473345$ & $-0.237917$ & $-0.131726$\tabularnewline
\hline 
$Y_{2}$ & $0.692487$ & $0.447724$ & $0.281942$ & $0.177417$\tabularnewline
\hline 
$Y_{3}$ & $0.276165$ & $0.276845$ & $0.369712$ & $0.567015$\tabularnewline
\hline 
$Z_{1}$ & $4.50251$ & $1.75438$ & $0.757442$ & $0.361606$\tabularnewline
\hline 
$Z_{2}$ & $-1.44450$ & $-1.01316$ & $-0.659927$ & $-0.400890$\tabularnewline
\hline 
$Z_{3}$ & $-0.189177$ & $0.218238$ & $0.720202$ & $1.25437$\tabularnewline
\hline 
\end{tabular}
\end{center}
\caption{\small Expansion coefficients $X$, $Y$, and $Z$ for $c\in\left\{0.1,0.2,...0.8,0.9\right\}$ necessary to generate simplified expansions for the time and $\theta$ averaged power $\Psi_{\text{non}}$ defined in \eqref{eq:Psi} for the non-dissipative regime. \label{tab:seriesnondiss}}
\end{table}

\begin{table}
\begin{center}
\begin{tabular}{|c|c|c|c|c|c|}
\hline 
 & $c=0.1$ & $0.2$ & $0.3$ & $0.4$ & $0.5$\tabularnewline
\hline 
\hline 
$X_{1}$ & $83.7324$ & $9.29576 $ & $2.29872 $ & $0.778738 $ & $0.316977$\tabularnewline
\hline 
$X_{2}$ & $-1.03116$ & $-0.755807$ & $-0.530064$ & $-0.353326$ & $-0.222088$\tabularnewline
\hline 
$X_{3}$ & $0.007595$ & $0.037407$ & $0.077634$ & $0.110910$ & $0.124928$\tabularnewline
\hline 
$Y_{1}$ & $-2580.36$ & $-153.661 $ & $-27.6906 $ & $-7.62614$ & $-2.61479$\tabularnewline
\hline 
$Y_{2}$ & $2.99587$ & $2.50703$ & $2.00887$ & $1.53179$ & $1.10309$\tabularnewline
\hline 
$Y_{3}$ & $0.190652$ & $0.380481$ & $0.393787$ & $0.250806$ & $0.022273$\tabularnewline
\hline 
$Z_{1}$ & $51991.8$ & $1589.38$ & $200.607$ & $44.2562$ & $12.9661$\tabularnewline
\hline 
$Z_{2}$ & $-2.37039$ & $-2.49343$ & $-2.46240$ & $-2.28560$ & $-1.98231$\tabularnewline
\hline 
$Z_{3}$ & $1.25910$ & $0.530613$ & $-0.37158$ & $-0.997227$ & $-1.21535$\tabularnewline
\hline 
\end{tabular}

\begin{tabular}{|c|c|c|c|c|}
\hline 
 & $c=0.6$ & $0.7$ & $0.8$ & $0.9$\tabularnewline
\hline 
\hline 
$X_{1}$ & $0.148287$ & $0.0780177$ & $0.045196$ & $0.028152$\tabularnewline
\hline 
$X_{2}$ & $-0.130337$ & $-0.070240$ & $-0.033345$ & $-0.011949$\tabularnewline
\hline 
$X_{3}$ & $0.115358$ & $0.086180$ & $0.047863$ & $0.014291$\tabularnewline
\hline 
$Y_{1}$ & $-1.03462$ & $-0.460474$ & $-0.228553$ & $-0.125221 $\tabularnewline
\hline 
$Y_{2}$ & $0.745189$ & $0.471083$ & $0.279817$ & $0.158215$\tabularnewline
\hline 
$Y_{3}$ & $-0.206710$ & $-0.358844$ & $-0.383868$ & $-0.263084$\tabularnewline
\hline 
$Z_{1}$ & $4.48153 $ & $1.73569$ & $0.741331$ & $0.349199$\tabularnewline
\hline 
$Z_{2}$ & $-1.58539$ & $-1.15663$ & $-0.767664$ & $-0.464386$\tabularnewline
\hline 
$Z_{3}$ & $-1.00824$ & $-0.470876$ & $0.235273$ & $0.968531$\tabularnewline
\hline 
\end{tabular}
\end{center}
\caption{\small Expansion coefficients $X$, $Y$, and $Z$ for $c\in\left\{0.1,0.2,...0.8,0.9\right\}$ necessary to generate simplified expansions for the time and $\theta$ averaged power $\Psi_{\text{diss}}$ defined in \eqref{eq:Psi} for the dissipative regime. \label{tab:seriesdiss}}
\end{table}

Given that we are considering the oblate case, which not only implies $I_{11}=I_{22}$ but also a single tumbling mode  similar to SAM rotation exists, as $\theta$ decreases from $90^{\circ}$ to $0^{\circ}$, we find that the general triaxial expression for time \eqref{eq:trelax} reduces to:
\begin{equation}
t_{\text{relax}}=-\left|\boldsymbol{J}\right|^{2}\left(\frac{1}{I_{22}}-\frac{1}{I_{33}}\right)\int_{90^{\circ}}^{0^{\circ}}\frac{\sin\theta\cos\theta\mathrm{d}\theta}{P_{\text{avg}}}\;\;. \label{eq:toblate}
\end{equation}
However, as argued above, $P_{\text{avg}}\approx\Psi$, which is independent of $\theta$, so that we can integrate for the characteristic dampening time exactly
\begin{equation}
t_{\text{relax}}\approx\frac{15\left(1-h^{2}\right)\left|\boldsymbol{J}\right|^{2}}{16\pi a^{5}h\left(h^{2}+1\right)\rho\Psi}\;\;,\label{eq:tdamp}
\end{equation}
where the relaxation time is given in seconds and we have used $I_{22}=4\pi\rho ha^5 \left(h^2+1\right)/15$ and $I_{33}=8\pi\rho ha^{5}/15$.

We illustrate the use of this estimate by applying it to an oblate \textquoteleft Oumuamua and Toutatis, using the same parameters that generated Fig. \ref{fig:PowerNorm}. In this case, the aspect ratios are $h=15/115\approx 0.13$ and $h=0.4909\approx 0.49$, so the closest approximation is therefore provided by expanding around $c=0.1$ and $c=0.5$. Substituting the expansion coefficients in the first column of Tables \ref{tab:seriesnondiss} and \ref{tab:seriesdiss} into \eqref{eq:Psi}, we obtain
\begin{align}
\Psi_{\text{non}}&=\frac{281525}{\eta}\;\;,\\
\Psi_{\text{diss}}&=1.44972\times 10^{9}\;,
\end{align}
which gives the characteristic dampening timescale using \eqref{eq:tdamp} as
\begin{align}
t_{\text{Oumuamua}}&\approx 1.54791\times 10^{-11}\eta\;\text{yrs}\;\;,\\
t_{\text{Toutatis}}&\approx 5.8754\times 10^{-8}\;\text{yrs}\;\;.
\end{align}

We comment on our result: Comparison of our analytic expression to numerical integration of \eqref{eq:toblate} gives a relative error no greater than $0.2\%$ varying $\eta$ from $10^{13}-10^{200}\;\text{Pa}.\text{s}$ for the case of `Oumuamua, and an error of $0.002\%$ for the case of Toutatis, suggesting that our averaging approximations and series expansions are valid. Furthermore, comparing our oblate estimate to the triaxial results presented in Section \ref{sec:Oumuamua}, we note that the relaxation time for `Oumuamua is between the timescales of LAM and SAM tumbling, whilst Toutatis is closer to its corresponding LAM result. To obtain the characteristic dampening time-scale obtained by \cite{fraser2018tumbling}, $\eta$ is now in the range of $10^{21}-10^{23}\;\text{Pa}.\text{s}$.

As a consequence of obtaining an analytic estimate for the relaxation timescale which features contributions from both rotation and self-gravitation, we can derive a criterion which indicates whether the effects of pre-stress can be neglected. We derive this from \eqref{eq:Psi} by demanding that, for a given trio of terms multiplied by a power of $(h-c)$, the latter two terms must be smaller than the first to ensure that gravity does not affect the relaxation dynamics. Explicitly, we find
\begin{equation}
|\boldsymbol{J}| \gg G^{1/2}\rho^{3/2}a^{5}\;,
\end{equation}
to ensure that self-gravitation can be correctly neglected for a body obeying a Maxwell rheology. We note that the parameters for Toutatis given in Table \ref{tab:Toutatis} do not satisfy this criterion, hence leading to the sizeable discrepancy between our results and those from previous works.

\section{Conclusion}

 In this paper, we studied the relaxation time-scale necessary to dampen the wobbling behaviour of a freely rotating inelastic ellipsoid.
 Following the ideas suggested in the recent work by \citet{frouard2017precession}, we determined the dissipation rate of the rotational kinetic energy by employing a method
 consistent with the field of continuum mechanics. The essence of this approach is to suppose that the rotator is obeying a general linear viscoelastic rheology, and to use this
 rheology to calculate the power dissipation from the distribution of the stresses and strains.

 In order to estimate the precession relaxation timescale, we developed a theory that first detailed the kinematics of a freely rotating object both in the LAM and SAM modes.
 Within the quasi-rigid and adiabatic approximations, we assumed that precession occurs on a timescale much shorter than relaxation, so the two processes could be
 addressed separately. The kinematics of the rotating body provided inertial forcing and were combined with the effects of self-gravitation to produce an elastic stress field for the rotator.

 At the next step, we obtained the viscoelastic stresses by the Correspondence Principle and the method of Laplace transforms. The latter was implemented due to the simple way of finding the inverse transform, which was a matter of obtaining the root of a polynomial and differentiating, rather than using integration or convolution operators as with other integral transforms. Having determined the viscoelastic stresses, we calculated the energy dissipation rate
 \footnote{~Aside from knowing the stress, it is also necessary to know the strain rate in order to write down the power. However, for linear rheologies, like the Maxwell example we did, this entire development can be just written in terms of the stress and stress rate.}
 and subsequently derived the relaxation time necessary for the rotator to dampen its precession to a specified final maximum wobbling angle.

 We then applied our formalism to the interstellar asteroid 1I/2017 (\textquoteleft Oumuamua) and the planet-orbit crossing 4179 Toutatis as examples of a close to non-dissipative body and a rotator which was highly dissipative. In both cases, weak deviatoric deformations were modeled by the Maxwell rheology whilst the volumetric deformations were taken to be elastic. In the non-dissipative regime, we numerically showed that the relaxation time-scales predicted in previous works were significant underestimates and corresponded to material parameters consistent with planetary mantles. Our model predicted that the relaxation time-scale was of the order $\,10^{23} - 10^{193}\;\text{yrs}\,$ when viscosities of very cold monoliths were used. This range of timescales exceeds the age of the universe so greatly that we may safely state that \textquoteleft Oumuamua's wobble has undergone no appreciable change during this asteroid's peregrination from where it was born.

Given that the mass density and second aspect ratio for \textquoteleft Oumuamua were unknown, we further investigated the effect of these two contributions on the process of precession relaxation. We have found that, generally, an increase of the mass density yields a decrease of the relaxation time, and further creates local maxima and minima in the dampening time as a function of the second-aspect ratio $\,h_2\,$, see Figure \ref{fig:densityOumuamua}. For LAM rotation, a local maximum and minimum are created with increasing mass density, whereas two local maxima and a minimum are created in the SAM mode.

We then shifted our focus to a highly energy dissipating body, Toutatis. Numerically, we found that employing a full mechanical treatment of precession relaxation produced a timescale of the order $\,10^{-8}\,\text{yr}\,$, which was significantly less than estimates produced by previous works. We determined that this discrepancy arose for two reasons: First, by employing viscosity rather than the empirical $Q$-factor and, second, by including the effect of self-gravitation which is no longer negligible for highly energy-dissipating bodies such as Toutatis. For the experimental viscosity reported, we found that the corresponding $Q$-factor was $\,10^{9}-10^{10}\,$ less than those posited by \cite{efroimsky2000inelastic}, whilst switching off gravity allowed us to obtain an estimate of the same order as that predicted by \cite{burns1973asteroid}. This example highlighted the significant differences that could arise from a full viscoelastic description of NPA relaxation and the consequence of omitting pre-stress, specifically for a highly dissipative rotator.
 
Having investigated the contributions of a triaxial geometry, we reduced our general theory to model a Maxwell rotator with an oblate geometry so that comparison could be made with previous works. We found that our work exhibited maximum power dissipation for both high and low wobbling angles, whereas previous works only showed a bias of high power dissipation for high wobbling angles. We argued that this was a consequence of including self-gravitation as we obtained similar results to those presented in \cite{frouard2017precession} when gravity was switched off. However, in our model, the power dissipation did not vanish when the body spun in its minimum rotational energy state as in previous works, however, we argued this result agreed with the Maxwell rheology imposed.

We then proceeded to derive simplified analytic estimates for the characteristic dampening time for both close to non-dissipating and highly dissipative bodies with Poisson ratios of $\,0.2\,$, which corresponded to cold bodies. By appealing to the symmetry of the power dissipation with respect to the wobbling angle and expanding in terms of a particular aspect ratio, we obtained a series of approximations which only exhibited relative errors no larger than $\,0.2\%\,$ for the case of an oblate `Oumuamua, and $\,0.002\%\,$ for an oblate Toutatis, when compared with numerical integration. Furthermore, we derived a criterion for when self-gravitation could be safely neglected and showed that the effects of pre-stress should be included when modeling the precession relaxation of Toutatis.

The current work is a promising step into mechanically describing the NPA tumbling relaxation of celestial bodies. There are numerous avenues of investigation that can be further pursued, such as the inclusion of more complex viscoelastic rheologies which feature secondary creeping behavior; a more realistic model for deformations in rock (see Appendix B for a brief discussion on the mathematics). However, with the significantly small relaxation timescale predicted for Toutatis, an important question to consider is whether the adiabatic approximation, allowing us to decouple the effects of rotation and deformation, remains valid and whether a significant error is produced by applying it to such highly deformable objects. In this case, one must then consider the full Euler equations for rotation, however, there do exist formalisms, such as that involving pseudo-elastic bodies, which may provide progress in answering such questions.

 \section*{Acknowledgments}

The author gratefully thanks the Japan Society for the Promotion of Science for funding this work, M. Efroimsky for initial motivation and gratuitous help in linking the mathematics to astronomical applications, Wade Henning and Julie Castillo-Rogez for helpful discussions regarding rheological parameters of cold rocks, Brian D. Warner and Petr Pravec for a highly valuable consultation on the Light Curve Database, T. G. Bollea for continued inspiration, and D. Paganin for very invaluable advice and support. The author dedicates this work \textit{ad maiorem Dei gloriam}.




\bibliographystyle{mnras}
\bibliography{RelaxingRotators} 




\appendix

\section{Coefficients for linearly elastic stresses}\label{sec:explicit}

We write the explicit forms for the elements of $\boldsymbol{S}^{\left(ij\right)}$ in ansatz \eqref{eq:ansatz}. By imposing the balance of angular momentum, the entries of the matrices satisfy $S_{mn}^{\left(ij\right)}=S_{nm}^{\left(ij\right)}$, for indices $m$ and $n$ taking the values $\left\{1,2,3\right\}$; in particular, the off-diagonal elements of a particular $\boldsymbol{S}^{\left(ij\right)}$ must be equal. By imposing the balance of linear momentum \eqref{eq:BLM} and the free-ends boundary condition \eqref{eq:BC}, we can solve $30$ of the entries in terms of the matrix elements of the constant matrix $S_{mn}^{\left(00\right)}$, which are
\begin{align}
S_{11}^{\left(11\right)} & =S_{11}^{\left(00\right)}, \qquad \qquad S_{12}^{\left(11\right)} =S_{12}^{\left(00\right)}, \qquad \qquad  S_{13}^{\left(11\right)} =S_{13}^{\left(00\right)}, \\
S_{22}^{\left(11\right)} & = \frac{h_{2}^{2}\left(h_{1}^{2}\left(2S_{11}^{\left(00\right)}-\rho\left(B_{11}+B_{11}-B_{33}\right)\right)+4S_{22}^{\left(00\right)}\right)-2S_{33}^{\left(00\right)}}{2h_{2}^{2}},\\
S_{23}^{\left(11\right)} & = 3S_{23}^{\left(00\right)}-\rho h_{1}^{2}h_{2}^{2}B_{23},\\
S_{33}^{\left(11\right)} & = \frac{h_{2}^{2}\left(h_{1}^{2}\left(2S_{11}^{\left(00\right)}-\rho\left(B_{11}-B_{22}+B_{33}\right)\right)-2S_{22}^{\left(00\right)}\right)+2S_{33}^{\left(00\right)}}{2},
\end{align}
\begin{align}
S_{12}^{\left(22\right)} & =S_{12}^{\left(00\right)}, \qquad \qquad S_{22}^{\left(22\right)} =S_{22}^{\left(00\right)}, \qquad \qquad  S_{23}^{\left(22\right)} =S_{23}^{\left(00\right)}, \\
S_{11}^{\left(22\right)} & = \frac{h_{2}^{2}S_{22}^{\left(00\right)}-S_{33}^{\left(00\right)}}{h_{1}^{2}h_{2}^{2}}+2S_{11}^{\left(00\right)}-\frac{\rho\left(B_{11}+B_{22}-B_{33}\right)}{2},\\
S_{13}^{\left(22\right)} & = 3S_{13}^{\left(00\right)}-\rho h_{1}^{2}h_{2}^{2}B_{13},\\
S_{33}^{\left(22\right)} & = \frac{h_{2}^{2}\left(h_{1}^{2}\left(2S_{11}^{\left(00\right)}+\rho\left(B_{22}+B_{33}-B_{11}\right)\right)-2S_{22}^{\left(00\right)}\right)}{2}-2S_{33}^{\left(00\right)},
\end{align}
\begin{align}
S_{13}^{\left(33\right)} & =S_{13}^{\left(00\right)}, \qquad \qquad S_{23}^{\left(33\right)} =S_{23}^{\left(00\right)}, \qquad \qquad  S_{33}^{\left(33\right)} =S_{33}^{\left(00\right)}, \\
S_{11}^{\left(33\right)} & = \frac{S_{33}^{\left(00\right)}-h_{2}^{2}S_{22}^{\left(00\right)}}{h_{1}^{2}h_{2}^{2}}+2S_{11}^{\left(00\right)}-\frac{\rho\left(B_{11}-B_{22}+B_{33}\right)}{2},\\
S_{12}^{\left(33\right)} & = 3S_{12}^{\left(00\right)}-\rho h_{1}^{2}B_{12},\\
S_{22}^{\left(33\right)} & = \frac{h_{2}^{2}\left(h_{1}^{2}\left(-2S_{11}^{\left(00\right)}+\rho\left(B_{11}-B_{22}+B_{33}\right)\right)+4S_{22}^{\left(00\right)}\right)+2S_{33}^{\left(0\right)}}{2h_{2}^{2}},
\end{align}
\begin{align}
S_{11}^{\left(12\right)} & =0, \qquad \qquad S_{22}^{\left(12\right)} = 0, \\
S_{33}^{\left(12\right)} & = 4h_{1}h_{2}^{2}S_{12}^{\left(00\right)}-2\rho h_{1}^{3}h_{2}^{2}B_{12},\\
S_{12}^{\left(12\right)} & = \frac{h_{2}^{2}\left(h_{1}^{2}\left(\rho\left(B_{11}+B_{22}-B_{33}\right)-2S_{11}^{\left(00\right)}\right)-2S_{22}^{\left(00\right)}\right)+2S_{33}^{\left(00\right)}}{2h_{1}h_{2}^{2}},\\
S_{13}^{\left(12\right)} & = -\frac{2S_{23}^{\left(00\right)}-\rho h_{1}^{2}h_{2}^{2}B_{23}}{h_{1}},\\
S_{23}^{\left(12\right)} & = -2h_{1}S_{13}^{\left(00\right)}+\rho h_{1}^{3}h_{2}^{2}B_{13},
\end{align}
\begin{align}
S_{11}^{\left(13\right)} & =0, \qquad \qquad S_{33}^{\left(13\right)} = 0, \\
S_{22}^{\left(13\right)} & = \frac{4h_{1}S_{13}^{\left(00\right)}-2\rho h_{1}^{3}h_{2}^{2}B_{13}}{h_{2}},\\
S_{12}^{\left(13\right)} & = -\frac{2S_{23}^{\left(00\right)}-\rho h_{1}^{2}h_{2}^{2}B_{23}}{h_{1}h_{2}},\\
S_{13}^{\left(13\right)} & = \frac{h_{2}^{2}\left(h_{1}^{2}\left(\rho\left(B_{11}-B_{22}+B_{33}\right)-2S_{11}^{\left(00\right)}\right)+2S_{22}^{\left(00\right)}\right)-2S_{33}^{\left(00\right)}}{2h_{1}h_{2}},\\
S_{23}^{\left(13\right)} & = -2h_{1}h_{2}S_{12}^{\left(00\right)}+\rho h_{1}^{3}h_{2}B_{12},
\end{align}
\begin{align}
S_{22}^{\left(23\right)} & =0, \qquad \qquad S_{33}^{\left(23\right)} = 0, \\
S_{11}^{\left(23\right)} & = \frac{4S_{23}^{\left(00\right)}-2\rho h_{1}^{2}h_{2}^{2}B_{23}}{h_{1}^{2}h_{2}},\\
S_{12}^{\left(23\right)} & = -\frac{2S_{13}^{\left(00\right)}-\rho h_{1}^{2}h_{2}^{2}B_{13}}{h_{2}},\\
S_{13}^{\left(23\right)} & = -2h_{2}S_{12}^{\left(00\right)}+\rho h_{1}^{2}h_{2}B_{12},\\
S_{23}^{\left(23\right)} & = \frac{h_{2}^{2}\left(h_{1}^{2}\left(2S_{11}^{\left(00\right)}+\rho\left(-B_{11}+B_{22}+B_{33}\right)\right)-2S_{22}^{\left(00\right)}\right)-2S_{33}^{\left(00\right)}}{2h_{2}}.\\
\end{align}

By further solving the constitutive relation \eqref{eq:BMequation}, we find the matrix entries of $\boldsymbol{S}^{\left(00\right)}$
\begin{align}
S_{11}^{\left(00\right)} & =\frac{\rho\left(f_{111}B_{11}-h_{1}^{2}f_{112}B_{22}-h_{1}^{2}h_{2}^{2}f_{113}B_{33}\right)}{g},\\
S_{22}^{\left(00\right)} & =\frac{\rho h_{1}^{2}\left(-f_{221}B_{11}+f_{222}B_{22}-h_{2}^{2}f_{223}B_{33}\right)}{g},\\
S_{33}^{\left(00\right)} & =\frac{\rho h_{1}^{2}h_{2}^{2}\left(f_{331}B_{11}+f_{332}B_{22}+f_{333}B_{33}\right)}{g},\\
S_{12}^{\left(00\right)}&=\frac{\rho h_{1}^{2}B_{12}\left(h_{1}^{2}h_{2}^{2}\left(2h_{2}^{2}+\nu+1\right)+h_{2}^{2}\left(\nu+1\right)+2(\nu+1)\right)}{2h_{2}^{2}\left(h_{1}^{2}\left(2h_{2}^{2}+\nu+1\right)+\nu+1\right)+6(\nu+1)},\\
S_{13}^{\left(00\right)}&=\frac{\rho h_{1}^{2}h_{2}^{2}B_{13}\left(h_{1}^{2}\left(h_{2}^{2}(\nu+1)+2\right)+\left(2h_{2}^{2}+1\right)(\nu+1)\right)}{2h_{1}^{2}\left(h_{2}^{2}(\nu+1)+2\right)+2\left(3h_{2}^{2}+1\right)(\nu+1)},\label{eq:example}\\
S_{23}^{\left(00\right)}&=\frac{\rho h_{1}^{2}h_{2}^{2} B_{23}\left(h_{1}^{2}\left(\nu+1\right)\left(2h_{1}^{2}h_{2}^{2}+h_{2}^{2}+1\right)+2\right)}{2h_{1}^{2}(\nu+1)\left(h_{2}^{2}\left(3h_{1}^{2}+1\right)+1\right)+4},
\end{align}
with $f$ and $g$ being purely functions of the Poisson ratio $\nu$, the former given by
\begin{align}
f_{111} & =3h_{1}^{8}h_{2}^{4}\left(\nu^{2}-1\right)\left(4h_{2}^{4}+h_{2}^{2}\left(\nu+3\right)+4\right)\nonumber\\
 & +h_{1}^{6}h_{2}^{2}\left(h_{2}^{2}+1\right)(\nu+1)\left(4h_{2}^{4}\left(\nu-2\right)\right.\nonumber\\
 & \left.+h_{2}^{2}\left(\nu\left(10\nu-7\right)-7\right)+4\left(\nu-2\right)\right)\nonumber\\
 & +h_{1}^{4}\left(8h_{2}^{8}\left(\nu^{2}-1\right)+h_{2}^{6}\left(\nu^{2}\left(7\nu+6\right)-9\nu-12\right)\right.\nonumber\\
 & +2h_{2}^{4}\left(\nu^{2}\left(11\nu-1\right)-7\nu-11\right)\nonumber\\
 & \left.+h_{2}^{2}\left(\nu^{2}\left(7\nu+6\right)-9\nu-12\right)+8\nu^{2}-8\right)\nonumber\\
 & +2h_{1}^{2}\left(h_{2}^{2}+1\right)\left(h_{2}^{4}\left(\nu^{2}-1\right)\left(2\nu+3\right)\right.\nonumber\\
 & \left.+h_{2}^{2}\left(\nu\left(\nu-3\right)-2\right)+\left(\nu^{2}-1\right)\left(2\nu+3\right)\right)\nonumber\\
 & +2\left(3h_{2}^{4}+2h_{2}^{2}+3\right)\left(\nu^{2}-1\right),
\end{align}
\begin{align}f_{112} & =h_{1}^{6}h_{2}^{4}(\nu^{2}-1\left(4h_{2}^{4}+3h_{2}^{2}\left(\nu+1\right)+2\right)\nonumber\\
 & +h_{1}^{4}h_{2}^{2}\left(4h_{2}^{6}\nu\left(\nu+1\right)+h_{2}^{4}\left(\nu^{2}(6\nu+13)-3\right)\right.\nonumber\\
 & \left.+h_{2}^{2}\left(\nu^{2}\left(8\nu+5\right)+4\nu-1\right)+2\left(4\nu^{2}+\nu-1\right)\right)\nonumber\\
 & +h_{1}^{2}\left(\nu h_{2}^{6}\left(5-\nu\right)\left(\nu+1\right)+h_{2}^{4}\left(2\nu^{2}\left(3\nu+4\right)+4\nu-2\right)\right.\nonumber\\
 & \left.+h_{2}^{2}\left(\nu^{2}\left(3\nu+4\right)+3\nu-2\right)+2\nu^{2}-2\right)\nonumber\\
 & -2\nu\left(\nu+1\right)\left(h_{2}^{4}\left(\nu-3\right)-2h_{2}^{2}\nu+\nu-1\right),
\end{align}
\begin{align}f_{113} & =h_{1}^{6}h_{2}^{2}\left(\nu^{2}-1\right)\left(2h_{2}^{4}+3h_{2}^{2}\left(\nu+1\right)+4\right)\nonumber\\
 & +h_{1}^{4}\left(2h_{2}^{6}\left(4\nu^{2}+\nu-1\right)+h_{2}^{4}\left(\nu^{2}\left(8\nu+5\right)+4\nu-1\right)\right.\nonumber\\
 & \left.+h_{2}^{2}\left(\nu^{2}\left(6\nu+13\right)-3\right)+4\nu\left(\nu+1\right)\right)\nonumber\\
 & +h_{1}^{2}\left(2h_{2}^{6}\left(\nu^{2}-1\right)+h_{2}^{4}\left(\nu^{2}\left(3\nu+4\right)+3\nu-2\right)\right.\nonumber\\
 & \left.+h_{2}^{2}\left(2\nu^{2}\left(3\nu+4\right)+4\nu-2\right)-\nu\left(\nu-5\right)\left(\nu+1\right)\right)\nonumber\\
 & +2\nu\left(h_{2}^{4}-\nu^{2}\left(h_{2}^{2}-1\right)^{2}+2\nu\left(h_{2}^{2}+1\right)+3\right),
\end{align}
\begin{align}f_{221} & =-2\left(\nu^{2}-1\right)\left(h_{1}^{2}\nu-1\right)+4h_{1}^{4}h_{2}^{8}\left(\nu+1\right)\left(h_{1}^{2}\nu+\nu-1\right)\nonumber\\
 & +h_{1}^{2}h_{2}^{6}\left(-h_{1}^{4}\nu\left(\nu-5\right)\left(\nu+1\right)+h_{1}^{2}\left(\nu^{2}\left(6\nu+13\right)-3\right)\right.\nonumber\\
 & \left.+3\left(\nu-1\right)\left(\nu+1\right)^{2}\right)+h_{2}^{2}\left(4h_{1}^{4}\nu^{2}\left(\nu+1\right)\right.\nonumber\\
 & \left.+h_{1}^{2}\left(\nu^{2}\left(3\nu+4\right)+3\nu-2\right)+2\left(4\nu^{2}+\nu-1\right)\right)\nonumber\\
 & +h_{2}^{4}\left(-2h_{1}^{6}\nu\left(\nu-3\right)\left(\nu+1\right)+h_{1}^{4}\left(2\nu^{2}\left(3\nu+4\right)+4\nu-2\right)\right.\nonumber\\
 & \left.+h_{1}^{2}\left(\nu^{2}\left(8\nu+5\right)+4\nu-1\right)+2\nu^{2}-2\right),
\end{align}
\begin{align}f_{222} & =2h_{1}^{8}h_{2}^{4}\left(\nu^{2}-1\right)\left(4h_{2}^{4}+h_{2}^{2}\left(2\nu+3\right)+3\right)\nonumber\\
 & +h_{1}^{6}h_{2}^{2}\left(4h_{2}^{6}\left(\nu-2\right)\left(\nu+1\right)+h_{2}^{4}\left(\nu^{2}\left(7\nu+6\right)-9\nu-12\right)\right.\nonumber\\
 & \left.+2h_{2}^{2}\left(2\nu^{2}\left(\nu+2\right)-5\nu-5\right)+4\nu^{2}-4\right)\nonumber\\
 & +h_{1}^{4}\left(12h_{2}^{8}\left(\nu^{2}-1\right)+h_{2}^{6}\left(\nu+1\right)\left(\nu\left(10\nu-3\right)-15\right)\right.\nonumber\\
 & +2h_{2}^{4}\left(\nu^{2}\left(11\nu-1\right)-7\nu-11\right)+2h_{2}^{2}\left(2\nu^{2}\left(\nu+2\right)-5\nu-5\right)\nonumber\\
 & \left.+6\nu^{2}-6\right)+h_{1}^{2}\left(3h_{2}^{6}\left(\nu^{2}-1\right)\left(\nu+3\right)\right.\nonumber\\
 & +h_{2}^{4}\left(\nu+1\right)\left(\nu\left(10\nu-3\right)-15\right)+h_{2}^{2}\left(\nu^{2}\left(7\nu+6\right)-9\nu-12\right)\nonumber\\
 & \left.+2(\nu^{2}-1)(2\nu+3)\right)\nonumber\\
&+4\left(\nu+1\right)\left(3h_{2}^{4}\left(\nu-1\right)+h_{2}^{2}\left(\nu-2\right)+2\left(\nu-1\right)\right),
\end{align}
\begin{align}f_{223} & =\nu\left(\nu+1\right)\left(4-h_{1}^{2}\left(2h_{1}^{2}\left(\nu-3\right)+\nu-5\right)\right)\nonumber\\
 & +2h_{1}^{4}h_{2}^{6}\left(h_{1}^{4}\left(\nu^{2}-1\right)+h_{1}^{2}\left(4\nu^{2}+\nu-1\right)+\nu^{2}-1\right)\nonumber\\
 & +h_{1}^{2}h_{2}^{4}\left(-2h_{1}^{6}\nu\left(\nu^{2}-1\right)+h_{1}^{4}\left(\nu^{2}\left(3\nu+4\right)+3\nu-2\right)\right.\nonumber\\
 & \left.+h_{1}^{2}\left(\nu^{2}\left(8\nu+5\right)+4\nu-1\right)+3\left(\nu-1\right)\left(\nu+1\right)^{2}\right)\nonumber\\
 & +h_{2}^{2}\left(4h_{1}^{6}\nu^{2}\left(\nu+1\right)+h_{1}^{4}\left(2\nu^{2}\left(3\nu+4\right)+4\nu-2\right)\right.\nonumber\\
 & \left.+h_{1}^{2}\left(\nu^{2}\left(6\nu+13\right)-3\right)+4\nu^{2}-4\right),
\end{align}
\begin{align}f_{331} & =h_{1}^{6}h_{2}^{2}\nu\left(\nu+1\right)\left(2h_{2}^{4}\left(\nu-3\right)+h_{2}^{2}\left(\nu-5\right)-4\right)\nonumber\\
 & -h_{1}^{4}\left(4h_{2}^{6}\nu^{2}\left(\nu+1\right)+h_{2}^{4}\left(2\nu^{2}\left(3\nu+4\right)+4\nu-2\right)\right.\nonumber\\
 & \left.+h_{2}^{2}\left(\nu^{2}\left(6\nu+13\right)-3\right)+4\nu^{2}-4\right)+h_{1}^{2}\left(2h_{2}^{6}\nu\left(\nu^{2}-1\right)\right.\nonumber\\
 & -h_{2}^{4}\left(\nu^{2}\left(3\nu+4\right)+3\nu-2\right)-h_{2}^{2}\left(\nu^{2}\left(8\nu+5\right)+4\nu-1\right)\nonumber\\
 & \left.-3\left(\nu-1\right)\left(\nu+1\right)^{2}\right)-2\left(h_{2}^{4}\left(\nu^{2}-1\right)\right.\nonumber\\
 & \left.+h_{2}^{2}\left(4\nu^{2}+\nu-1\right)+\nu^{2}-1\right),
\end{align}
\begin{align}f_{332} & =2h_{1}^{8}h_{2}^{4}\left(\nu^{2}-1\right)\left(h_{2}^{2}\nu-1\right)-h_{1}^{6}h_{2}^{2}\left(4h_{2}^{4}\nu^{2}\left(\nu+1\right)\right.\nonumber\\
 & \left.+h_{2}^{2}\left(\nu^{2}\left(3\nu+4\right)+3\nu-2\right)+2\left(4\nu^{2}+\nu-1\right)\right)\nonumber\\
 & +h_{1}^{4}\left(2h_{2}^{6}\nu\left(\nu-3\right)\left(\nu+1\right)-2h_{2}^{4}\left(\nu^{2}\left(3\nu+4\right)+2\nu-1\right)\right.\nonumber\\
 & \left.-h_{2}^{2}\left(\nu^{2}\left(8\nu+5\right)+4\nu-1\right)-2\nu^{2}+2\right)\nonumber\\
 & +h_{1}^{2}\left(h_{2}^{4}\nu\left(\nu-5\right)\left(\nu+1\right)+h_{2}^{2}\left(3-\nu^{2}\left(6\nu+13\right)\right)\right.\nonumber\\
 & \left.-3\left(\nu-1\right)\left(\nu+1\right)^{2}\right)-4\nu\left(h_{2}^{2}\left(\nu+1\right)+\nu\right)+4,
\end{align}
\begin{align}f_{333} & =2h_{1}^{8}h_{2}^{4}\left(\nu^{2}-1\right)\left(3h_{2}^{4}+h_{2}^{2}\left(2\nu+3\right)+4\right)\nonumber\\
 & +h_{1}^{6}h_{2}^{2}\left(4h_{2}^{6}\left(\nu^{2}-1\right)+2h_{2}^{4}\left(2\nu^{2}\left(\nu+2\right)-5\nu-5\right)\right.\nonumber\\
 & \left.+h_{2}^{2}\left(\nu^{2}\left(7\nu+6\right)-9\nu-12\right)+4\left(\nu-2\right)\left(\nu+1\right)\right)\nonumber\\
 & +h_{1}^{4}\left(6h_{2}^{8}\left(\nu^{2}-1\right)+2h_{2}^{6}\left(2\nu^{2}\left(\nu+2\right)-5\nu-5\right)\right.\nonumber\\
 & +2h_{2}^{4}\left(\nu^{2}\left(11\nu-1\right)-7\nu-11\right)\nonumber\\
 & +h_{2}^{2}\left(\nu+1\right)\left(\nu\left(10\nu-3\right)-15\right)\left.+12\left(\nu^{2}-1\right)\right)\nonumber\\
 & +h_{1}^{2}\left(2h_{2}^{6}\left(\nu^{2}-1\right)\left(2\nu+3\right)\right.\nonumber\\
 & +h_{2}^{4}\left(\nu^{2}\left(7\nu+6\right)-9\nu-12\right)\nonumber\\
 & +h_{2}^{2}\left(\nu+1\right)\left(\nu\left(10\nu-3\right)-15\right)\left.+3\left(\nu^{2}-1\right)\left(\nu+3\right)\right)\nonumber\\
 & +4\left(\nu+1\right)\left(2h_{2}^{4}\left(\nu-1\right)\right.\left.+h_{2}^{2}\left(\nu-2\right)+3\left(\nu-1\right)\right),
\end{align}
whilst the latter is found as
\begin{align}g & =6h_{1}^{8}h_{2}^{4}\left(\nu^{2}-1\right)\left(4h_{2}^{4}+h_{2}^{2}\left(\nu+3\right)+4\right)\nonumber\\
 & +2h_{1}^{6}h_{2}^{2}\left(h_{2}^{2}+1\right)\left(\nu^{2}-1\right)\left(8h_{2}^{4}+h_{2}^{2}\left(13\nu+7\right)+8\right)\nonumber\\
 & +2h_{1}^{4}\left(\nu-1\right)\left(12h_{2}^{8}\left(\nu+1\right)+h_{2}^{6}\left(\nu+1\right)\left(13\nu+15\right)\right.\nonumber\\
 & \left.+h_{2}^{4}\left(38\nu^{2}+50\nu+32\right)+h_{2}^{2}\left(\nu+1\right)\left(13\nu+15\right)+12\left(\nu+1\right)\right)\nonumber\\
 & +2h_{1}^{2}\left(h_{2}^{2}+1\right)\left(\nu^{2}-1\right)\left(3h_{2}^{4}\left(\nu+3\right)+\right.\nonumber\\
 & \left.+2h_{2}^{2}\left(5\nu+3\right)+3\left(\nu+3\right)\right)+8\left(\nu^{2}-1\right)\left(3h_{2}^{4}+2h_{2}^{2}+3\right).\label{eq:g}
\end{align}

\section{Proof of the simplified form for the Inverse Laplace Transform under linear rheologies}\label{sec:proof}

\begin{figure}
 \begin{center}
    \includegraphics[width=0.3\textwidth]{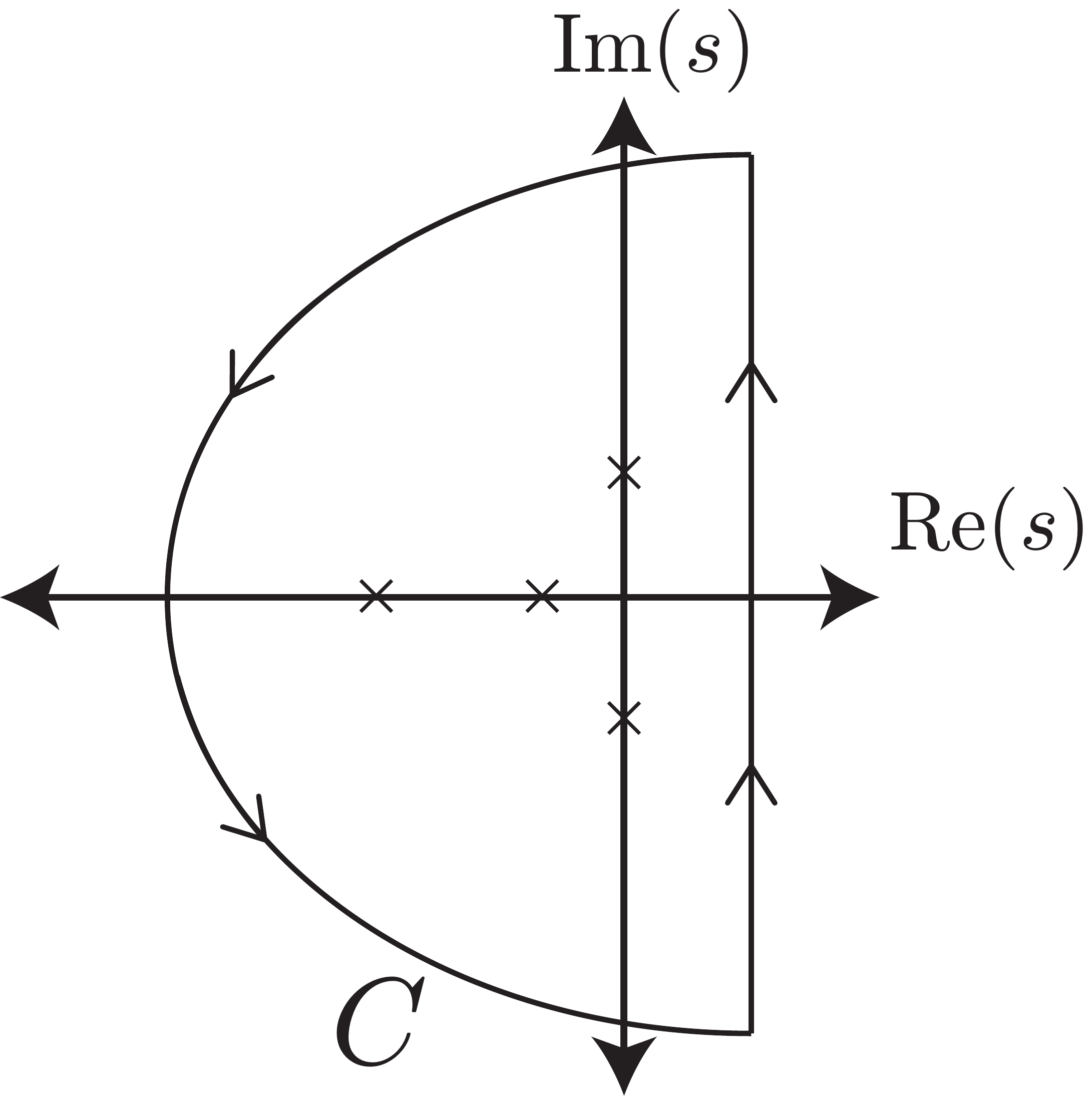}
 \end{center}
 \caption{Integration in complex frequency space to determine the inverse Laplace transform. The real parameter $\beta$ in the Bromwich integral \eqref{eq:GENinverseLap} must be chosen so that all singularities of the function $e^{st}\widehat{\alpha}\left(s\right)$ (shown as crosses) are to the left of the vertical contour. For linear viscoelastic rheologies, one is then allowed to create an enclosed contour in the complex plane that contains all the poles, thereby reducing the problem of inverse transformation to finding the corresponding residues of the singularities. \label{fig:contour}}
 \end{figure}

We provide a more detailed discussion of the mathematics leading to the reduction of the inverse Laplace transform to \eqref{eq:inverseLap} for linear viscoelastic rheologies. The most general method of the inverse Laplace transform involves calculation of the Bromwich integral \citep{bromwich1917normal}
\begin{equation}
\alpha\left(t\right)\equiv\mathcal{L}^{-1}\left\{\widehat{\alpha}\left(s\right)\right\}=\lim_{\xi\rightarrow\infty}\frac{1}{2\pi \mathrm{i}}\intop_{\beta-\mathrm{i}\xi}^{\beta+\mathrm{i}\xi}e^{st}\widehat{\alpha}\left(s\right)\mathrm{d}s\;\;, \label{eq:GENinverseLap}
\end{equation}
where $\beta\in\mathbb{R}$ is a parameter chosen so that all the singularities of $\widehat{\alpha}\left(s\right)$, taken to occur at $s=s_{0}$, satisfy $\mathrm{Re}\left(s_{0}\right)<\beta$; namely, that the singularities occur to the left of the vertical contour integral in complex frequency space (see Fig. \ref{fig:contour}).

Given that the focus of the current work is on linear viscoelastic rheologies, the transformed operators $\widehat{P}_{1}$, $\widehat{U}_{1}$, $\widehat{P}_{2}$, and $\widehat{U}_{2}$ feature only integer powers of the Laplace variable $s$, and therefore only pole singularities exist in the integrand. This aspect allows us to create a closed, right-hand oriented contour in the complex frequency space, so that
\begin{equation}
\oint e^{st}\widehat{\alpha}\left(s\right)\mathrm{d}s = \lim_{\xi\rightarrow\infty}\intop_{\beta-\mathrm{i}\xi}^{\beta+\mathrm{i}\xi}e^{st}\widehat{\alpha}\left(s\right)\mathrm{d}s +\intop_{C}e^{st}\widehat{\alpha}\left(s\right)\mathrm{d}s\;\;,
\end{equation}
where $C$ is a left semi-circular contour with the center of the circle lying at $s=\beta$ (see Fig. \ref{fig:contour}).

 To evaluate the integral over $C$, we use the contour parameterisation $s=\beta+r e^{\mathrm{i}\psi}$ for $\psi\in\left[\pi/2,3\pi/2\right]$ in the limit of $r\rightarrow\infty$,
 given that $\xi\rightarrow\infty$. In this limit, the transformed generalised Poisson ratio \eqref{eq:genPRs} becomes
\begin{equation}
\widehat{\nu}_{VE}\sim\frac{p_{1}^{\left(m_{1}\right)}u_{2}^{\left(n_{2}\right)}s^{m_{1}+n_{2}}-p_{2}^{\left(m_{2}\right)}u_{1}^{\left(n_{1}\right)}s^{m_{2}+n_{1}}}{p_{1}^{\left(m_{1}\right)}u_{2}^{\left(n_{2}\right)}s^{m_{1}+n_{2}}+2p_{2}^{\left(m_{2}\right)}u_{1}^{\left(n_{1}\right)}s^{m_{2}+n_{1}}}\;\;,
\end{equation}
which gives three possibilities:
\begin{itemize}
\item $m_{1}+n_{2}<m_{2}+n_{1}$, so that $\widehat{\nu}_{VE}\sim -1/2$,
\item $m_{1}+n_{2}>m_{2}+n_{1}$, yielding the behaviour $\widehat{\nu}_{VE}\sim 1$,
\item and $m_{1}+n_{2}=m_{2}+n_{1}$, so that $\widehat{\nu}_{VE}\sim\frac{\left(p_{1}^{\left(m_{1}\right)}u_{2}^{\left(n_{2}\right)}-p_{2}^{\left(m_{2}\right)}u_{1}^{\left(n_{1}\right)}\right)}{\left(p_{1}^{\left(m_{1}\right)}u_{2}^{\left(n_{2}\right)}+2p_{2}^{\left(m_{2}\right)}u_{1}^{\left(n_{1}\right)}\right)}$.
\end{itemize}
The second and third cases cause the integral over $C$ to be non-trivial: The second case results in a singularity, given that this is a zero of $g$ found in \eqref{eq:g}, which results in poles for the explicit solutions for the entries of $\boldsymbol{S}^{\left(00\right)}$, whilst the third can lead similarly cause $g$ to vanish for particular values of $p_{1}^{\left(m_{1}\right)}$, $p_{2}^{\left(m_{2}\right)}$, $u_{1}^{\left(n_{1}\right)}$, and $u_{2}^{\left(n_{2}\right)}$.

Restricting our attention to the first and third case, for constants which do not cause stress singularities, the fictitious stress remains bounded. Furthermore, we note that, for the present work, $\widehat{\alpha}\left(s\right)$ can be decomposed as:
\begin{equation}
\widehat{\alpha}\left(s\right)=\mathcal{X}\left(\widehat{\nu}_{VE}\right)\mathcal{L}\left\{B_{ij}\right\}\;\;,
\end{equation}
whereby the function $\mathcal{X}$ is independent of $s$ in the limit $r\rightarrow\infty$, as $\,\hat{\nu}_{VE}\,$ asymptotically reduces to the constants discussed above.

As a direct consequence, we are guaranteed a finite inverse Laplace transform given that the Laplace transform of $B_{ij}$ exists, leading to the nice reduction:
\begin{equation}
\oint e^{st}\widehat{\alpha}\left(s\right)\mathrm{d}s = \lim_{\xi\rightarrow\infty}\intop_{\beta-\mathrm{i}\xi}^{\beta+\mathrm{i}\xi}e^{st}\widehat{\alpha}\left(s\right)\mathrm{d}s\;\;.
\end{equation}

We compute the closed contour integral using Cauchy's Residue Theorem, which states that the location and nature of the singularities enclosed in the contour determine the computed value of the integral \citep{mitrinovic1984cauchy}
\begin{equation}
2\pi\mathrm{i}\sum \mathrm{Res}\left[e^{st}\widehat{\alpha}\left(s\right)\right]= \lim_{\xi\rightarrow\infty}\intop_{\beta-\mathrm{i}\xi}^{\beta+\mathrm{i}\xi}e^{st}\widehat{\alpha}\left(s\right)\mathrm{d}s\;\;,
\end{equation}
so that, by the relation between the Bromwich integral and the inverse Laplace transform, we obtain the desired simplified form
\begin{equation}
\mathcal{L}^{-1}\left\{\widehat{\alpha}\left(s\right)\right\}=\sum\mathrm{Res}\left[e^{st}\widehat{\alpha}\left(s\right)\right]\;\;.
\end{equation}

We make some closing remarks on the emphasis that the powers of $s$ in the deviatoric and volumetric operators be integer powers. A natural question to consider is what occurs when fractional powers of $s$, which predict secondary relaxation timescales like the Andrade model \citep{efroimsky2012bodily}, are included. The answer is that such singularities are now branch points, which introduce notions of multi-valued functions. The simple complex integration contour in Fig. \ref{fig:contour} that allowed us to derive the simplified form of the inverse Laplace transform breaks down; it is no longer guaranteed that the integrand has the same value along this entire contour. Therefore, the integration contour must now be deformed to avoid crossing into a Riemann manifold where the function has a different value, which hence results in integral contributions that no longer necessarily vanish, as for the case above.


 \label{lastpage}
 \end{document}